\documentclass[a4paper,fleqn,usenatbib]{mnras}

\usepackage[T1]{fontenc}
\usepackage{ae,aecompl}

\usepackage{graphicx}	
\usepackage{amsmath}	
\usepackage{amssymb}	

\DeclareMathOperator{\n}{\text{n}}
\DeclareMathOperator{\p}{\text{p}}
\DeclareMathOperator{\df}{\mathrm{d}\!}

\title[Relativistic simulations of pulsar glitches]{Global numerical simulations of the rise of vortex-\\mediated pulsar glitches in full general relativity}

\author[A. Sourie et al.]{
A. Sourie,$^{1}$\thanks{E-mail: aurelien.sourie@obspm.fr} 
N. Chamel,$^{2}$\thanks{E-mail: nchamel@ulb.ac.be}
J. Novak$^{1}$\thanks{E-mail: jerome.novak@obspm.fr}
and M. Oertel$^{1}$\thanks{E-mail: micaela.oertel@obspm.fr}
\\
$^{1}$Laboratoire Univers et Th\'eories, Observatoire de Paris, PSL Research University, CNRS,
Universit\'e Paris Diderot, \\ Sorbonne Paris Cit\'e, 5 place Jules Janssen, 92195 Meudon, France\\
$^{2}$Institut d'Astronomie et d'Astrophysique, CP-226,
Universit\'e Libre de Bruxelles, 1050 Brussels, Belgium
}

\date{Accepted XXX. Received YYY; in original form ZZZ}

\pubyear{2016}

\begin{document}
\label{firstpage}
\pagerange{\pageref{firstpage}--\pageref{lastpage}}
\maketitle

\begin{abstract}
In this paper, we study in detail the role of general relativity on the global dynamics of giant pulsar glitches as exemplified by Vela. For this purpose, we carry out numerical simulations of the spin up triggered by the sudden unpinning of superfluid vortices. In particular, we compute the exchange of angular momentum between the core neutron superfluid and the rest of the star within a two-fluid model including both (non-dissipative) entrainment effects and (dissipative) mutual friction forces. Our simulations are based on a quasi-stationary approach using realistic equations of state (EoSs) following \cite{sourie2016numerical}. We show that the evolution of the angular velocities of both fluids can be accurately described by an exponential law. The associated characteristic rise time $\tau_{\text{r}}$, which can be precisely computed from stationary configurations only, has a form similar to that obtained in the Newtonian limit. However, general relativity changes the structure of the star and leads to additional couplings between the fluids due to frame-dragging effects. As a 
consequence, general relativity can have a large impact on the actual value of $\tau_{\text{r}}$: the errors incurred by using Newtonian gravity are thus found to be as large as $\sim 40 \%$ for the models considered. Values of the rise time are calculated for Vela and compared with current observational limits. Finally, we study the amount of gravitational waves emitted during a glitch. Simple expressions are obtained for the corresponding characteristic amplitudes and frequencies. The detectability of glitches through gravitational wave observatories is briefly discussed.
\end{abstract}




\section{Introduction}
\label{intro}

Pulsars are very compact stars rotating rapidly with exceptionally stable periods spanning from $\sim$ 1.4 milliseconds to a few seconds. Nevertheless, some pulsars exhibit sudden increases in their observed angular velocity $\Omega$, with relative amplitude $\Delta \Omega/ \Omega$ ranging between $\sim 10^{-11}$ and $\sim 10^{-5}$ \citep{wong2001observations, espinoza2011study}. These spin-up events, known as \emph{glitches}, are usually followed by a slow relaxation on time scales up to months or years and are sometimes accompanied by abrupt changes in the pulsar spin-down rate, $\Delta\dot{\Omega}/\dot{\Omega}\sim 10^{-4} - 10^{-2}$ (we use a dot to denote time derivative). Presently, 472 glitches have been detected in 165 pulsars\footnote{http://www.jb.man.ac.uk/pulsar/glitches.html.} \citep{espinoza2011study}, with angular velocities ranging from 0.09 Hz to 327 Hz (see, e.g., the ATNF Pulsar Database\footnote{http://www.atnf.csiro.au/research/pulsar/psrcat.}; \citet{manchester2005australia}). At least 
two distinct glitching behaviours have 
been identified \citep{espinoza2011study, yu2013detection}: (i) quasi-periodic giant glitches with a very narrow spread in size around $\Delta\Omega/\Omega \sim 10^{-6}$, and (ii) smaller glitches of various sizes at random intervals of time. The most emblematic pulsar of the first kind is Vela (PSR~B0833--45) with a rotation frequency $f=\Omega /(2\pi) \simeq 11.19$~Hz \citep{dodson2007two} corresponding to a period $P=1/f\simeq 89$~ms. Since its discovery in 1969, 19 glitches have been detected so far every $\sim$ 2-3 years~\citep{espinoza2011study}. The second type of glitching pulsars is exemplified by the Crab (PSR~B0531+21) with a rotation frequency $f \simeq  29.95$~Hz ($P\simeq 33$~ms). 

Since the first detections of glitches \citep{radhakrishnan1969detection, reichley1969observed}, different mechanisms have been proposed to explain these events (see, e.g., the review by \citet{haskell2015models}). A glitch is nowadays commonly thought as the manifestation of an internal process, except possibly for highly magnetised neutron stars for which some evidence of magnetospheric activity have been found (e.g. \citet{archibald2013anti, keith2013connection, antonopoulou2015unusual}). The interior of neutron stars can thus be probed using observations of pulsar glitches. 

Glitches were first suggested to arise from crustquakes \citep{ruderman1969neutron, baym1971neutron}. Following this idea, the presence of a solid crust (which crystallized when the star 
was young and rapidly rotating) prevents readjustments of the stellar shape, as the star spins down due to electromagnetic emission. Crustal stresses thus build up, until the crust cracks and the star suddenly adopts a more spherical shape. The resulting reduction of the moment of inertia leads the pulsar to spin up, assuming conservation of angular momentum. This scenario can account for small glitches, such as those exhibited by the Crab pulsar. However, as pointed out by  \citet{ruderman1969neutron}, this mechanism fails to predict the occurrence frequency of giant glitches, as observed in the Vela pulsar.

\begin{figure}
\center 
\includegraphics[width = \columnwidth]{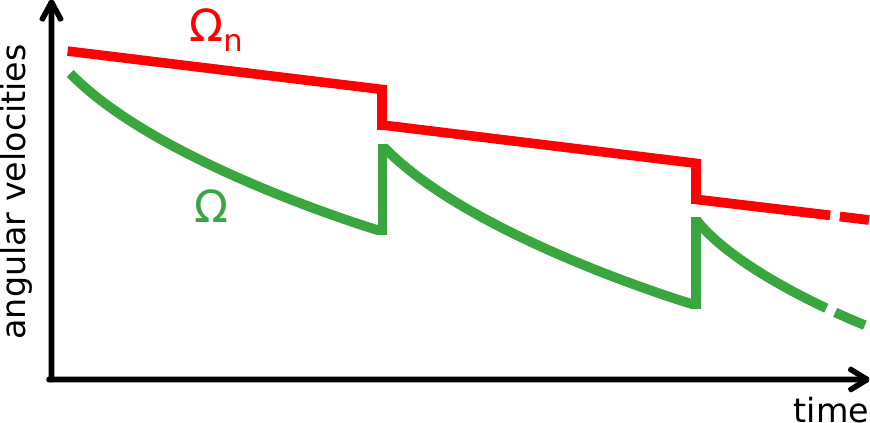}
\caption{Schematic time evolution of the observed angular velocity $\Omega$ of the pulsar and of the angular velocity $\Omega_{\n}$ of the interior neutron superfluid 
over two glitch events.}
\label{fig:evol}
\end{figure}

Giant glitches are generally thought to be the manifestation of superfluid matter inside neutron stars, as suggested by the very long time scales observed during post-glitch relaxations \citep{baym1969superfluidity}. From theoretical calculations, the interior of a neutron star is expected to contain an isotropic neutron superfluid in the inner crust, an anisotropic neutron superfluid in the outer core, and possibly other superfluid species in the inner core (see, e.g., \cite{page2013stellar}). In a seminal work, \citet{anderson1975pulsar} proposed that glitches themselves could be triggered by the sudden unpinning of neutron superfluid vortices. The idea is the following. It is well-known from laboratory experiments (see, e.g., \citet{yarmchuk1979observation, abo-shaeer2001observation, zwierlein2005vortices}) that a superfluid can only rotate by forming an array of quantized vortices, each carrying a quantum $\hbar$ of angular momentum, where $\hbar$ is the Planck-Dirac constant. The neutron superfluid 
present 
in the core and the inner crust of neutron stars is thus expected to be threaded by a huge number of vortex lines, with a mean surface density given by 
\begin{equation}
\label{surfdens_vortex}
n_{v} \, (\text{cm}^{-2}) = \frac{4 m_{\n} \Omega_{\n}}{h} \simeq\frac{10^7}{P(\text{ms})}\, ,
\end{equation} where $h=2\pi \hbar$ is the Planck constant, $m_{\n}$ is the neutron rest mass and the coarse-grained averaged angular velocity $\Omega_{\n}$ of the neutron superfluid is approximated by that of the star \citep{ginzburg1965}. The neutron superfluid is supposed to be weakly coupled to the rest of the star by so-called mutual friction forces arising from the dissipative forces acting on individual vortices~\citep{alpar1984rapid}. Due to the spin down of the star induced by the electromagnetic torque, vortices tend to move away from the rotation axis. The key assumption of vortex-mediated glitch theories is that vortices can pin to nuclear clusters in the inner crust~\citep{anderson1975pulsar} 
and/or to quantized magnetic flux tubes in the core if protons form a type II superconductor \citep{baym1969superfluidity,sauls1989superfluidity, ruderman1998neutron}. In such case, the neutron superfluid is decoupled from the rest of the star, and can rotate more rapidly, as schematically illustrated on Fig.~\ref{fig:evol}. The lag $\delta\Omega = \Omega_{\n} - \Omega >0 $ induces a Magnus force on the vortices. The larger the lag, the stronger the force. For some critical value $\delta \Omega_0$ of the lag, vortices will suddenly unpin, the superfluid will spin down and, by conservation of angular momentum, the rest of the star will spin up leading to the observed glitch (see Fig.~\ref{fig:evol}).
During the subsequent relaxation, these vortices are thought to progressively repin and a lag can grow anew. This vortex-mediated scenario is supported by laboratory experiments \citep{tsakadze1980properties} and the ability of the vortex creep model to reproduce the post-glitch relaxations in different pulsars \citep{alpar1984vortex, alpar1984vortexII, alpar1993postglitch, alpar1996postglitch, gugercinoglu2014vortex}.

It is noteworthy to mention that the two mechanisms described above are not necessarily independent. 
Indeed, starquakes can be induced by the presence of superfluids in neutron star interiors, whether superfluid vortices are pinned~\citep{ruderman1991neutron} or not~\citep{carter2000centrifugal, chamel2006effect}. In turn, sudden motions of vortices can be triggered by quakes~\citep{ruderman1991neutron,chau1993correlated, alpar1996postglitch,eichler2010}. 

Although mesoscopic studies of large collections ($\sim10^2-10^4$) of vortices provide useful insight (e.g. \cite{warszawski2011gross, warszawski2013knock}), simulating pulsar glitches requires to follow the dynamics of all the superfluid vortices contained in the star. Given their huge number, $\sim 10^{17}$ for Vela, the overall transfer of angular momentum between the neutron superfluid and the rest of the star can be studied using a smooth-averaged hydrodynamic approach, still involving microscopic parameters determined by the local dynamics of individual vortices (see, e.g., \cite{bulgac2013strength} and references therein). Whereas the general relativistic framework for describing starquakes was developed a long time ago~\citep{carter1975relativistic}, the general relativistic formulation of the vortex-mediated glitch model is more recent~\citep{langlois1998differential}. As a matter of fact, most global numerical simulations of pulsar glitches have been performed within the Newtonian framework~(e.g.,
 \cite{larson2002simulations, peralta2006transitions, sidery2010dynamics,haskell2012modelling}). Recently, \cite{seveso2012effect} and \cite{antonelli2016axially} have developed a non-relativistic hydrodynamic model for describing the different stages of the glitch phenomenon based on the static structure of the neutron star computed in general relativity. However, general relativity could also play an important role for the global dynamics of glitches. Furthermore, general relativity is essential to determine the amount of gravitational waves associated with glitch events. Observations of gravitational waves are of particular interest since they could potentially provide additional information on the glitch phenomenon (see, e.g.,  \cite{stopnitzky2014gravitational, haskell2015models} and references therein).

In this paper, we present global numerical simulations of vortex-mediated pulsar glitches. We focus on the spin-up stage regardless of the glitch triggering mechanism. On the other hand, we study the glitch dynamics in full general relativity.
We also derive the associated gravitational wave characteristic amplitudes and frequencies using  the standard quadrupole formula. The paper is organized as follows. We start by presenting, in Section~\ref{typical}, the different assumptions on which our model is based. In Section~\ref{sec:evol_eq}, we introduce the evolution equations governing the transfer of angular momentum in the interior of a pulsar during a glitch. 
Results of stationary rotating configurations are discussed in Section~\ref{stat_conf}. In Section~\ref{numerical}, we detail the numerical procedure underlying our simulations. Results for the glitch rise time are presented and discussed. We study the emission of gravitational waves in Section~\ref{gws}. Finally, we conclude in Section~\ref{conclusion}.

\section{Model assumptions}
\label{typical}

\subsection{Quasi-stationary approach}
\label{quasi_stat}
The glitch phenomenon can be decomposed into distinct stages (\textit{i.e.} the pre-glitch evolution, the spin up, and the post-glitch relaxation), which can be 
modelled separately in view of the different associated time scales suggesting different physical mechanisms. Focusing on the sudden spin up of the pulsar after the catastrophic unpinning of vortices,  stellar dynamics are essentially governed by the mutual friction force 
between the superfluid and the rest of the star. This force acts on a characteristic time scale corresponding to the glitch \textit{rise} time $\tau_{\text{r}}$, which has not been fully observationally resolved yet. The most stringent observational constraint on $\tau_{\text{r}}$ comes from the 2000 and 2004 Vela glitch timing data: $\tau_{\text{r}} < 30 - 40$~s \citep{dodson2002high,dodson2007two}.

It is interesting to compare $\tau_{\text{r}}$ with the typical time
$\tau_{\text{h}}$ for the star to go back to hydrodynamic equilibrium
once being driven out of it by a change in its rotation
rate.  Sometimes referred to as \textit{hydrodynamic} time scale, this time is roughly given by \citep{shapiro1983black}
\begin{equation}
  \tau_{\text{h}} \sim (G \bar{\rho})^{-\frac{1}{2}} \simeq 0.1  \left( \frac{R}{12 \text{ km} } \right)^{\frac{3}{2}} \left( \frac{M}{1.4\ \text{M}_{\odot}}\right)^{-\frac{1}{2}} \text{ms},  
\end{equation}
where $G$ denotes the gravitational constant, $M$ is the neutron-star mass, $R$ the stellar radius, and $\bar{\rho}\sim 3 M/(4 \pi R^3)$ is the average density of the star. The hydrodynamic time $\tau_{\text{h}}$ represents the time for a sound wave with speed $c_s$ to propagate throughout a star of radius $R$, \textit{i.e.} $\tau_{\text{h}} \sim R / c_s$ \citep{epstein1988acoustic}. 

In the following, we shall assume that $ \tau_{\text{r}} \gg \tau_{\text{h}}$, as suggested by previous studies~\citep{haskell2012modelling}, 
so that the dynamical evolution of the pulsar can be reasonably well described by a sequence of quasi-stationary equilibrium configurations. 

\subsection{Two-component model}

Due to the magnetic field, the electrically charged particles inside neutron stars are strongly coupled and essentially co-rotate with the crust and the magnetosphere 
at the observed angular velocity $\Omega$~\citep{glampedakis2011magnetohydro}. We do not account for other effects of the magnetic field on the global dynamics of the star, 
which could be important for the most strongly magnetised neutron stars~\citep{bocquet1995rotating, chatterjee2015consistent}, but can be safely ignored for the ordinary pulsars considered here. 

The simplest model of pulsars thus consists of at least two distinct dynamical components \citep{baym1969}: (i) a plasma of charged particles (electrons, nuclei in the crust, and protons in the core), and (ii) a neutron superfluid extending in the whole core. Because of (non-dissipative) mutual neutron-proton entrainment effects according to which the momenta are misaligned with the corresponding velocities~\citep{andreev1976three}, neutron superfluid vortices in the core of a neutron star carry a fractional magnetic quantum flux~\citep{sedrakian1980mechanism}. Electrons scattering off the magnetic field of the vortex lines was shown to induce a strong coupling between the core superfluid and the crust~\citep{alpar1984rapid}. For this reason, only the neutron superfluid permeating the inner crust of the star has been generally thought to be responsible for giant glitches. This scenario was also supported by the analysis of the glitch data, which suggested that the superfluid represents only a few percent of 
the angular momentum reservoir of the star~\citep{alpar1993postglitch, datta1993implications,link1999pulsar}. On the other hand, this model has been recently challenged~\citep{chamel2006effect,andersson2012pulsar,
chamel2013crustal,delsate2016giant} by the realization that despite the absence of viscous drag the crust can still resist the flow of the neutron superfluid due to Bragg scattering~\citep{chamel2004phd,carter2005entrainment,chamel2005band,chamel2012neutron}.
It has been argued that crustal entrainment could be much weaker assuming that the superfluid coherence length is much smaller than the size of clusters~\citep{martin2016superfluid}. However, as recognized by these authors, this condition is generally not satisfied. 
Even if crustal entrainment is ignored, the analyses of the 2007 glitch detected in PSR~J1119$-$6127, as well as of the 2010 glitch in PSR~B2334$+$61 indicate that the crust is not  enough~\citep{yuan2010very,alpar2011largest,akbal2015peculiar}. These recent studies suggest that the core superfluid plays a more important role than previously thought. In particular, the core superfluid could be decoupled from the rest of the star due to the pinning of vortices to quantized magnetic flux tubes assuming protons form a type II superconductor~\citep{gugercinoglu2014vortex}. 

In this work, we thus focus on the dynamics of the superfluid neutron star core within a two-fluid model: a neutron superfluid coupled to 
the ``normal'' fluid made of protons and electrons (simply referred to as ``protons" in the following). Quantities related to the two fluids will be labelled by indices ``n'' and ``p'' respectively. Note that, since in our model we do not consider the stellar crust, we do not account for any crust-core coupling mechanisms, such as Ekman pumping. These couplings could still have a strong impact on the glitch dynamics, especially during the post-glitch relaxation \citep{vanEysden2010pulsar, haskell2015models}.

\subsection{Spacetime symmetries}

Our glitch simulations are based on the general relativistic equilibrium configurations of rotating superfluid neutron stars computed by~\cite{sourie2016numerical}. In this section, we recall the main assumptions on the spacetime symmetries and the metric. 

The star is supposed to be axisymmetric and stationarily rotating. Neglecting the very small non-circular motion of the neutron superfluid due to the radial displacement of the vortices during the glitch \citep{langlois1998differential}, the two fluids are further assumed to rotate around a common axis with possibly different rotation rates. The spacetime is thus stationary, axisymmetric, circular and asymptotically flat. While the angular velocity $\Omega_{\p}=\Omega$ of the charged components can be reasonably assumed to be uniform, the angular velocity $\Omega_{\n}$ of the neutron superfluid may vary throughout the star. We circumvent this difficulty by considering that both fluids are rigidly rotating as in the model of \cite{sidery2010dynamics} in Newtonian gravity. In this case, $\Omega_{\n}$ and $\Omega_{\p}$ are to be understood as the angular velocities averaged over the whole star.  
More details on general relativistic equilibrium configurations of rotating superfluid neutron stars can be found in \cite{prix2005relativistic, sourie2016numerical}.

\subsection{Chemical composition and equation of state}
\label{chemical}

The dominant electroweak processes governing the composition of a neutron star are the direct (DU) and modified (MU) Urca beta processes \citep{yakovlev2001neutrino}. Within the assumption of rigid-body rotation with a non-vanishing lag $\delta\Omega$, beta equilibrium can only possibly be achieved on the rotational axis of the star~\citep{andersson2001slowly}. Assuming corotation ($\delta \Omega=0$) and ignoring superfluidity, the relaxation times towards beta equilibrium are roughly given by \citep{yakovlev2001neutrino}
\begin{align}
\label{rel_beta1}
  \tau_{\beta}^{(DU)} &\simeq 20 \left(\frac{T}{10^9 \ \text{K}} \right)^{-4} \ \text{s} \\
  \label{rel_beta2}
  \tau_{\beta}^{(MU)} &\simeq \left(\frac{T}{10^9 \ \text{K}}
  \right)^{-6} \ \text{months}
\end{align}
where $T$ represents the interior temperature of the star. For glitching pulsars, whose characteristic ``ages" $\tau_{\text{sd}} = \Omega/(2|\dot{\Omega}|) > 10^3$ years \citep{espinoza2011study} correspond to temperatures below $\sim 10^9$ K \citep{gnedin2001thermal}, Eqs.~(\ref{rel_beta1}) and (\ref{rel_beta2}) lead to time scales of the order of a few tens of seconds and a month for the DU and MU processes respectively. Differential rotation could in principle change the chemical equilibrium~\citep{langlois1998differential}, however the resulting effects are presumably negligible in view of the very small lags $\delta \Omega \ll \Omega$. On the contrary, superfluidity can strongly reduce the rates of beta processes \citep{villain2005non}, making these time scales even longer. Therefore, beta equilibrium may not be achieved during the spin up. 
  
The DU and MU rates being poorly known, we assume that $\tau_{\beta}$ still remains much shorter than the interglitch time so that the star is in beta equilibrium on the rotational axis at the beginning of a glitch. During the glitch rise, in order to estimate the error induced by our lack of knowledge, we restrict to the two non-dissipative limiting cases:
\begin{enumerate}
\item $ \tau_{\beta} \gg \tau_\text{r}$, \textit{i.e.} no reaction takes place during the glitch. The baryon masses of the two fluids $M^B_{\n}$ and $M^B_{\p}$ thus remain separately constant, which leads stellar matter on the rotational axis to be (slightly) out of beta equilibrium. 
  \vspace{0.2 cm}
\item $ \tau_{\beta} \ll \tau_\text{r}$, \textit{i.e.} the stellar matter on the rotational axis goes back instantaneously to beta equilibrium so that the chemical potentials satisfy $\mu^{\n} =  \mu^{\p}$, where $\mu^{\n}$ ($\mu^{\p}$) represents the neutron (proton) chemical potential. In our simulations, it is sufficient to impose this condition at the center of the star, \textit{i.e.} $\mu^{\n}_c = \mu^{\p}_c$, as discussed by \citet{prix2005relativistic}. 
Only the total baryon mass $M^B =  M^B_{\n}+ M^B_{\p}$ is conserved in this case. 
\end{enumerate}

We adopt the same equations of state (EoSs) as in \citet{sourie2016numerical}. These EoSs referred to as DDH and DDH$\delta$ were calculated from density-dependent relativistic mean-field models, including $\sigma$, $\omega$, $\rho$ mesons for the former and in addition the $\delta$ meson for the latter \citep{typel1999relativistic, avancini2009nuclear}. They were adapted to a system of two fluids at zero temperature coupled by entrainment for arbitrary compositions.

 \section{Evolution equations}
 \label{sec:evol_eq}
 
 \subsection{Angular momentum transfer}
\label{transfer}

Let $J_{\n}$ and $J_{\p}$ be the neutron superfluid and proton fluid angular momenta respectively (see~\cite{langlois1998differential} and \cite{sourie2016numerical} for definitions and expressions). Neglecting any external torque, the dynamics of these two fluids during the spin up is simply governed by the following equations 
  \begin{equation}
\label{evol_eq}
  \left\{
      \begin{aligned}
        \dot{J}_{\n} &=\ +\ \Gamma_{\text{mf}}, \\
        \dot{J}_{\p} &=\ -\ \Gamma_{\text{mf}}, \\
      \end{aligned}
    \right.
\end{equation}
where $\Gamma_{\text{mf}}$ stands for the mutual friction torque and overdot for time derivative.

\citet{langlois1998differential} derived a covariant expression for the (relativistic) mutual friction torque $\Gamma_{\text{mf}}$, 
considering straight vortices parallel to the rotation axis and arranged on a regular array. The dynamical evolution of the superfluid might potentially lead to the 
formation of a vortex tangle \citep{peralta2006transitions, andersson2007superfluid}. However, the onset of superfluid turbulence remains highly speculative, and therefore  
we shall not consider this possibility here. In \cite{langlois1998differential}, the  motion of unpinned vortex lines was assumed to be simply determined from the interplay between a Magnus force due to the neutron fluid and a dissipative drag force caused by the proton fluid.

Our numerical approach is based on the 3+1 formalism, in which the spacetime is foliated by a family $\left( \Sigma_t \right)_{t \in \mathbb{R}}$ of space-like hypersurfaces (see, e.g., \cite{gourgoulhon20123+}). Neglecting the small contribution of the non-circular motion of the vortices, the 3+1 expression of the mutual friction torque derived by~\cite{langlois1998differential} reads
 \begin{equation}
\Gamma_{\text{mf}} =  \displaystyle\int_{\Sigma_{t}} \mathcal{B} \
\Gamma_{\n} n_{\n} \varpi_{\n} \chi_{\perp}^2   \df^{\, 3}\! \Sigma
\times \left(\Omega_{\p} - \Omega_{\n}\right),  
\label{mom}
\end{equation}
in the absence of any dissipation related to chemical reactions, see Eqs.~(72) and (89) of \cite{langlois1998differential}. In this expression, $\df^{\, 3}\! \Sigma$  denotes the volume element on the hypersurface $\Sigma_{\text{t}}$ and $n_{\n}$ is the density of the neutron fluid in its rest frame. The term $\Gamma_{\n}$ stands for the Lorentz factor of the neutrons with respect to the so-called Eulerian observer or zero-angular-momentum observer (ZAMO), whose 4-velocity corresponds to the unit future-oriented (time-like) vector normal to $\Sigma_{t}$ (e.g. \citet{gourgoulhon20123+}). The  macroscopic neutron vorticity $\varpi_{\n}$ reads (we use Greek letters for spacetime indices)
\begin{equation}
\label{def_vort}
\varpi_{\n} = \sqrt{\frac{\varpi_{\mu\nu}\varpi^{\mu\nu}}{2}}~,
\end{equation}
where the vorticity 2-form $\varpi_{\mu\nu}$ is defined by
\begin{equation}
\label{def_w}
\varpi_{\mu\nu} = \nabla_{\mu}p^{\n}_{\nu} -  \nabla_{\nu}p^{\n}_{\mu}~, 
\end{equation}
$p^{\n}_{\mu}$ denoting the conjugate superfluid momentum. We note here that, on length scales smaller than the intervortex separation  $d_v$, typically of the order of $d_v\sim n_v^{-1/2}\simeq 10^{-3}$~cm (see Eq.~(\ref{surfdens_vortex})), $\varpi_{\mu\nu}$ strictly vanishes because $p^{\n}_{\mu}$ should be locally proportional to the gradient of a quantum scalar phase. Nevertheless, on the large scales we are interested in here, the neutron vorticity 2-form is non-vanishing, as well as its corresponding scalar amplitude $\varpi_{\n}$. Moreover, the geometric quantity $\chi_{\perp}^2$ is defined by 
 \begin{equation}
\label{def_hperp}
\chi_{\perp}^2 = \chi_{\perp}^{\alpha} \chi_{\perp \alpha}, \ \ \ \  \chi_{\perp}^{\alpha}= \perp^{\alpha}_{\beta}\chi^{\beta}, 
\end{equation}
where $\chi^{\alpha}$ is the Killing vector associated with axisymmetry and $\perp_{\alpha\beta}$ is the projection tensor orthogonal to the 2-dimensional string-type world sheets representing the vortex cores, see \cite{langlois1998differential} for details. 

Finally, the mutual friction parameter $\mathcal{B}$ characterizes the efficiency of the angular momentum transfer through mutual friction. This parameter is given by (see e.g., \cite{langlois1998differential,carter2001relativistic})
\begin{equation}
\label{drag_to_lift}
\mathcal{B} = \frac{\mathcal{R}}{1+ \mathcal{R}^2}
\end{equation}
as a function of the positive dimensionless drag-to-lift ratio $\mathcal{R}$. Since the dissipative processes contributing to mutual friction are not the same in different stellar regions \citep{alpar1984rapid,jones1990rotation, jones1992rotation, epstein1992vortex, sedrakian1995superfluid, haskell2014new}, $\mathcal{R}$ is expected to vary throughout the star. However, the values of this coefficient remain very 
uncertain. Microscopic estimates differ by many orders of magnitude. 
Given the current lack of knowledge on the microscopic origin of the mutual friction force, and since we are interested in global models of neutron stars, we introduce the averaged coefficient 
 \begin{equation}
\bar{\mathcal{B}} = \frac{ \displaystyle\int_{\Sigma_{t}} \mathcal{B}\ \Gamma_{\n} n_{\n} \varpi_{\n} \chi_{\perp}^2   \df^{\, 3}\! \Sigma
}{ \displaystyle\int_{\Sigma_{t}} \
\Gamma_{\n} n_{\n} \varpi_{\n} \chi_{\perp}^2   \df^{\, 3}\! \Sigma
}, 
\label{Bbar}
\end{equation}
that we consider as a free input parameter in our numerical simulations. Although $\mathcal{R}$ is likely to vary in time during the glitch event (due to changes of the vortex velocity or the repinning of some vortices, for instance), $ \bar{\mathcal{B}}$ is assumed to be time-independent for simplicity. The mutual friction torque (\ref{mom}) thus becomes 
\begin{equation}
\label{mom2}
\Gamma_{\text{mf}} = - \bar{\mathcal{B}} \displaystyle\int_{\Sigma_{t}}  \ \Gamma_{\n} n_{\n} \varpi_{\n} \chi_{\perp}^2  \df^{\, 3}\! \Sigma \times \delta \Omega, 
\end{equation}
where $\delta \Omega = \Omega_{\n} - \Omega_{\p}$ is the lag between the fluids. As shown in Appendix~\ref{torque_newt}, the Newtonian limit of Eq.~(\ref{mom2}) is in perfect agreement with the expression given by~\citet{sidery2010dynamics}, see their Eq.~(58).

To describe any transfer of angular momentum, it is convenient to introduce the partial moments of inertia 
\begin{equation}
\label{def_I}
I_{X\hspace*{-0.05 cm}X} = \left(\frac{\partial J_{X}}{\partial \Omega_{X}}\right)_{\Omega_{Y}} \ \ \text{and} \ \ \  I_{X\hspace*{-0.05 cm}Y} = \left(\frac{\partial J_{X}}{\partial \Omega_{Y}}\right)_{\Omega_{X}},
\end{equation}
where the two different capital letters $X$ and $Y$ refer to protons or neutrons. Depending on the assumption on the chemical composition (see Sec.~\ref{chemical}), these derivatives are taken either for fixed partial baryon masses $M^B_{\n}$ and $M^B_{\p}$ (case~i) or for a constant total baryon mass $M^B$ with identical chemical potentials at the center of the star (case~ii). It is possible to show that $I_{\n\hspace*{-0.05 cm}\p} =  I_{\p\hspace*{-0.05 cm}\n} $, see Eq.~(3.10) from \cite{carter1975application}. Furthermore, the partial moments of inertia should obey the following conditions 
\begin{equation}
\label{condition}
I_{\n\hspace*{-0.05 cm}\n} >0, \ \  I_{\p\hspace*{-0.05 cm}\p} >0 \ \ \& \ \ I_{\n\hspace*{-0.05 cm}\n}I_{\p\hspace*{-0.05 cm}\p} > I_{\n\hspace*{-0.05 cm}\p}^{\ 2},
\end{equation}
see Appendix~\ref{app:constraints_stab} for more details. We also define the neutron and proton moments of inertia, $\hat{I}_{\n}$ and $\hat{I}_{\p}$, as
\begin{equation}
\hat{I}_{\n} = I_{\n\hspace*{-0.05 cm}\n} + I_{\n\hspace*{-0.05 cm}\p} \ \ \ \text{and} \ \ \  \hat{I}_{\p} = I_{\p\hspace*{-0.05 cm}\p} + I_{\n\hspace*{-0.05 cm}\p}, 
\end{equation}
and the total moment of inertia $\hat{I}$  by $\hat{I} = \hat{I}_{\n} + \hat{I}_{\p}$. 
We note here that these definitions are more general than the moments of inertia discussed in Sec.~II-D of \citet{sourie2016numerical} in the limiting case of corotating fluids. However, the two definitions coincide in the slow-rotation approximation.

Given the previous definitions, the time derivatives of the angular momenta can be expressed as
 \begin{equation}
\label{der_J}
  \left\{
      \begin{aligned}
\dot{J}_{\n} &= I_{\n\hspace*{-0.05 cm}\n} \dot{\Omega}_{\n} + I_{\n\hspace*{-0.05 cm}\p}\dot{\Omega}_{\p}, \\
\dot{J}_{\p} &= I_{\n\hspace*{-0.05 cm}\p}\dot{\Omega}_{\n} + I_{\p\hspace*{-0.05 cm}\p}\dot{\Omega}_{\p}. \\
      \end{aligned}
    \right.
\end{equation}
Using the expression (\ref{mom2}) of the mutual friction torque, the angular momentum transfer (\ref{evol_eq}) reads
\begin{equation}
\label{evol_eq2}
  \left\{
      \begin{aligned}
        \dot{\Omega}_{\n} &= - \ \frac{\hat{I}_{\p}}{I_{\n\hspace*{-0.05 cm}\n}I_{\p\hspace*{-0.05 cm}\p} - I_{\n\hspace*{-0.05 cm}\p}^{\ 2}} \times  \bar{\mathcal{B}} \ \kappa \  \delta \Omega, \\
\dot{\Omega}_{\p} &= + \ \frac{\hat{I}_{\n}}{I_{\n\hspace*{-0.05 cm}\n}I_{\p\hspace*{-0.05 cm}\p} - I_{\n\hspace*{-0.05 cm}\p}^{\ 2}} \times  \bar{\mathcal{B}} \ \kappa \  \delta \Omega, \\
      \end{aligned}
    \right.
\end{equation}
where we have introduced the quantity $\kappa$ defined by
\begin{equation}
\label{def_kappa}
\kappa = \displaystyle\int_{\Sigma_{t}}  \ \Gamma_{\n} n_{\n} \varpi_{\n} \chi_{\perp}^2  \df^{\, 3}\! \Sigma. 
\end{equation}
It should be noticed that, in view of the conditions~(\ref{condition}), the denominator appearing in Eq.~(\ref{evol_eq2}) never vanishes. The time evolution of the lag $\delta \Omega = \Omega_{\n} - \Omega_{\p}$ is thus governed by the simple equation
\begin{equation}
\label{der_lag}
\dfrac{\delta{\dot{\Omega}}}{\delta\Omega}= - \ \frac{\hat{I}}{I_{\n\hspace*{-0.05 cm}\n}I_{\p\hspace*{-0.05 cm}\p} - I_{\n\hspace*{-0.05 cm}\p}^{\ 2 }} \times  \bar{\mathcal{B}} \ \kappa.
\end{equation}

\subsection{Analytical estimate for the spin-up time scale}
\label{estimate}

We now focus on deriving an approximate analytical formula for the spin-up time scale. Recalling that $\Delta \Omega / \Omega \ll 1$, where $\Delta \Omega$ represents the variation in the pulsar rotation rate during the spin up, it is a reasonable approximation to neglect the change in the different quantities appearing in the right-hand side of (\ref{der_lag}). Starting from an initial lag $\delta \Omega_0$ at the beginning of the spin up (see Sec.~\ref{IC}), the lag therefore approximately evolves as 
\begin{equation}
\delta \Omega (t) \approx \delta \Omega_0 \times \exp(-t/\tau_{\text{r}}),
\label{lag_evol}
\end{equation}
where we have introduced the characteristic time scale 
\begin{equation}
\label{tau_r1}
 \tau_{\text{r}} = \frac{\hat{I}_{\n}\hat{I}_{\p} - \hat{I}I_{\n\hspace*{-0.05 cm}\p} }{\hat{I}  \bar{\mathcal{B}} \kappa}.
\end{equation}
The time evolution of the two angular velocities is given by
\begin{equation}
 \label{time_evol}
   \begin{array}{rcl}
 \Omega_{\n}(t) &\approx&  \Omega_{\n}^0 + \dfrac{\hat{I}_{\p}}{\hat{I}}\times \delta \Omega_0\times \left( \exp(-t/\tau_{\text{r}}) - 1 \right), \vspace*{0.15 cm} \\ 
 \Omega_{\p}(t) &\approx&  \Omega_{\p}^0 - \dfrac{\hat{I}_{\n}}{\hat{I}}\times \delta \Omega_0\times \left( \exp(-t/\tau_{\text{r}}) - 1 \right), 
 \end{array}
\end{equation}
where $\Omega_{\n}^0$ and $\Omega_{\p}^0$ are the fluid rotation rates when the glitch is triggered. 

By analogy with the Newtonian limit (\ref{kappa_newt}), we introduce the quantity $\zeta$ through the relation 
\begin{equation}
\kappa = 2\zeta\hat{I}_{\n}\Omega_{\n}.
\label{zeta}
\end{equation}
Still, it should be remarked that general relativistic corrections are not only included in $\zeta$ but are also partially contained in $\hat{I}_{\n}$. Using (\ref{zeta}), the general relativistic rise time (\ref{tau_r1}) now reads
\begin{equation}
\label{tau_r}
\tau_{\text{r}} = \dfrac{\hat{I}_{\p}}{\hat{I}}\times\dfrac{1}{2\zeta\bar{\mathcal{B}} \Omega_{\n}} \times \left(1- \dfrac{I_{\n\hspace*{-0.05 cm}\p}\hat{I} }{\hat{I}_{\n}\hat{I}_{\p}} \right).
\end{equation}

Considering slowly rotating stars, for which $\Omega_{\n}, \Omega_{\p} \ll \Omega_{K}$ where the Keplerian limit $\Omega_{K}$ is the maximum angular velocity above which mass-shedding occurs at the equator, the Newtonian limit of (\ref{tau_r}) is found to be in perfect agreement with results from \citet{sidery2010dynamics}, see Appendix~\ref{rise_newt}. Note that this approximation is quite reasonable, given the rotation frequencies of observed glitching pulsars and the estimated values of the Keplerian frequency $\Omega_{K}/(2\pi)$, which is of the order of  $\sim 900$ to $1800$~Hz for neutron stars with masses larger than 1.4~M$_{\odot}$ (e.g., \cite{haensel2009keplerian,fantina2013neutron,haensel2016rotating}).

The analytical expression (\ref{tau_r}) of the spin-up time scale calls for several remarks. First, since the radial velocity of the vortex lines increases with the mutual friction parameter (see Eq.~(36) of \citet{carter2001relativistic} in the Newtonian case and Eq.~(85) of \citet{langlois1998differential} in the relativistic framework), the larger $\bar{\mathcal{B}}$ is, the faster is the transfer of angular momentum, consistently with (\ref{tau_r}).
As pointed out by \cite{carter2001relativistic}, the parameter $\bar{\mathcal{B}}$ can not take any arbitrary value: according to (\ref{drag_to_lift}), we have $\bar{\mathcal{B}}\leq 1/2$. 
This implies the following lower bound for the spin-up time scale:
\begin{equation}
\label{lower_bound}
\tau_{\text{r}}  \geqslant \dfrac{\hat{I}_{\p}}{\hat{I}}\times\dfrac{1}{\zeta  \Omega_{\n}} \times \left(1- \dfrac{I_{\n\hspace*{-0.05 cm}\p}\hat{I} }{\hat{I}_{\n}\hat{I}_{\p}} \right).
\end{equation}
which is only reached for $\mathcal{R}=1$. Furthermore, in the slow-rotation approximation, $\tau_{\text{r}}$ is also found to be  inversely proportional to $\Omega_{\n}$,
 which can be interpreted from the fact that the mutual friction torque is proportional to the surface density of vortex lines (see Eq.~(65) of \citet{andersson2006mutual} in the Newtonian context) through the superfluid vorticity $\varpi_{\n}$ (\ref{def_vort}), which in turn is roughly proportional to $\Omega_{\n}$, see Eq.~(\ref{surfdens_vortex}). The roles of $\zeta$ and $I_{\n\hspace*{-0.05 cm}\p}$ will be studied in details in the following section.

\section{Stationary~rotating~configurations}
\label{stat_conf}

In this section, we present numerical results concerning the superfluid vorticity and the couplings between the fluids, which are both playing an important role in the angular momentum transfer during a glitch event. These quantities can be directly obtained from the equilibrium configurations computed by \cite{sourie2016numerical}.

\subsection{Superfluid vorticity}
\label{sup_vorti}

\begin{figure*}
\includegraphics[width = 0.49\textwidth]{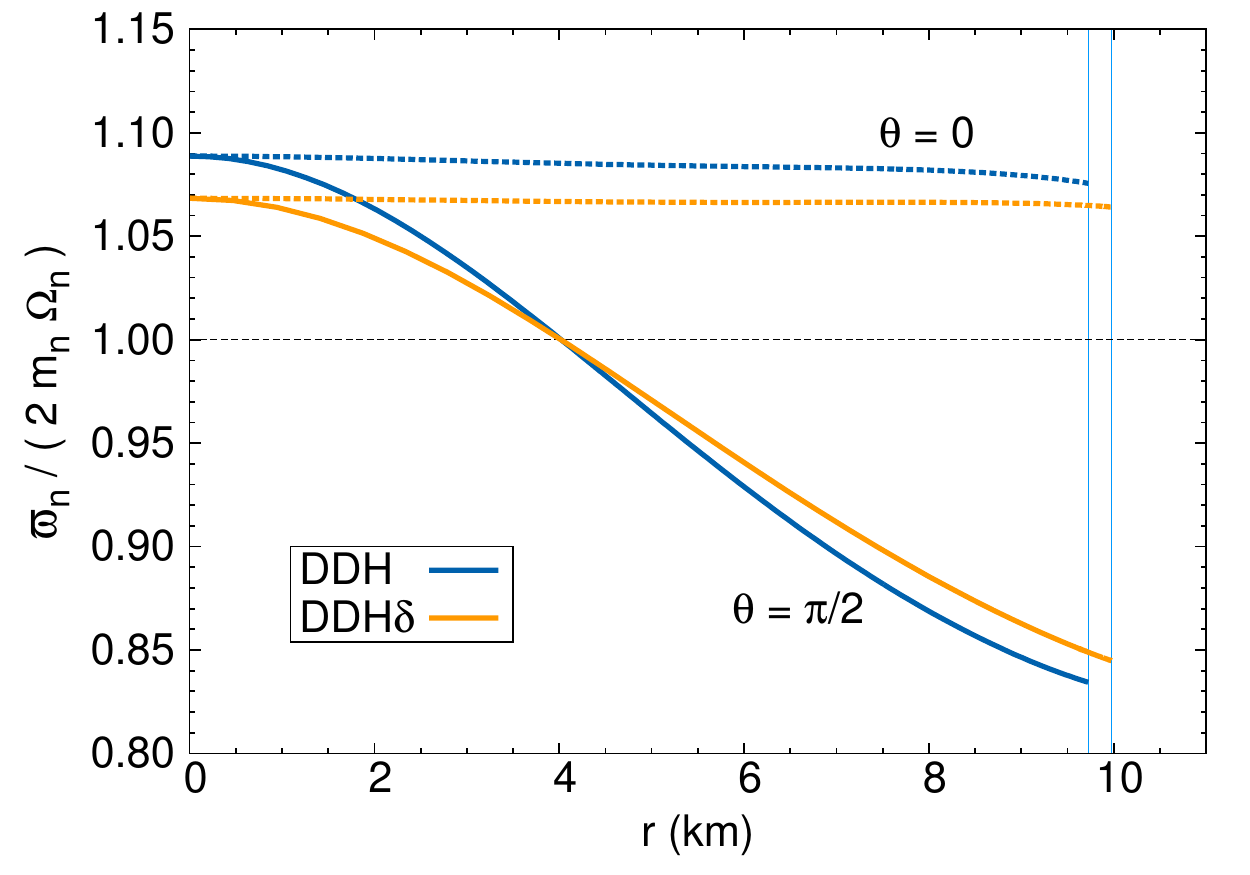}
\includegraphics[width = 0.49\textwidth]{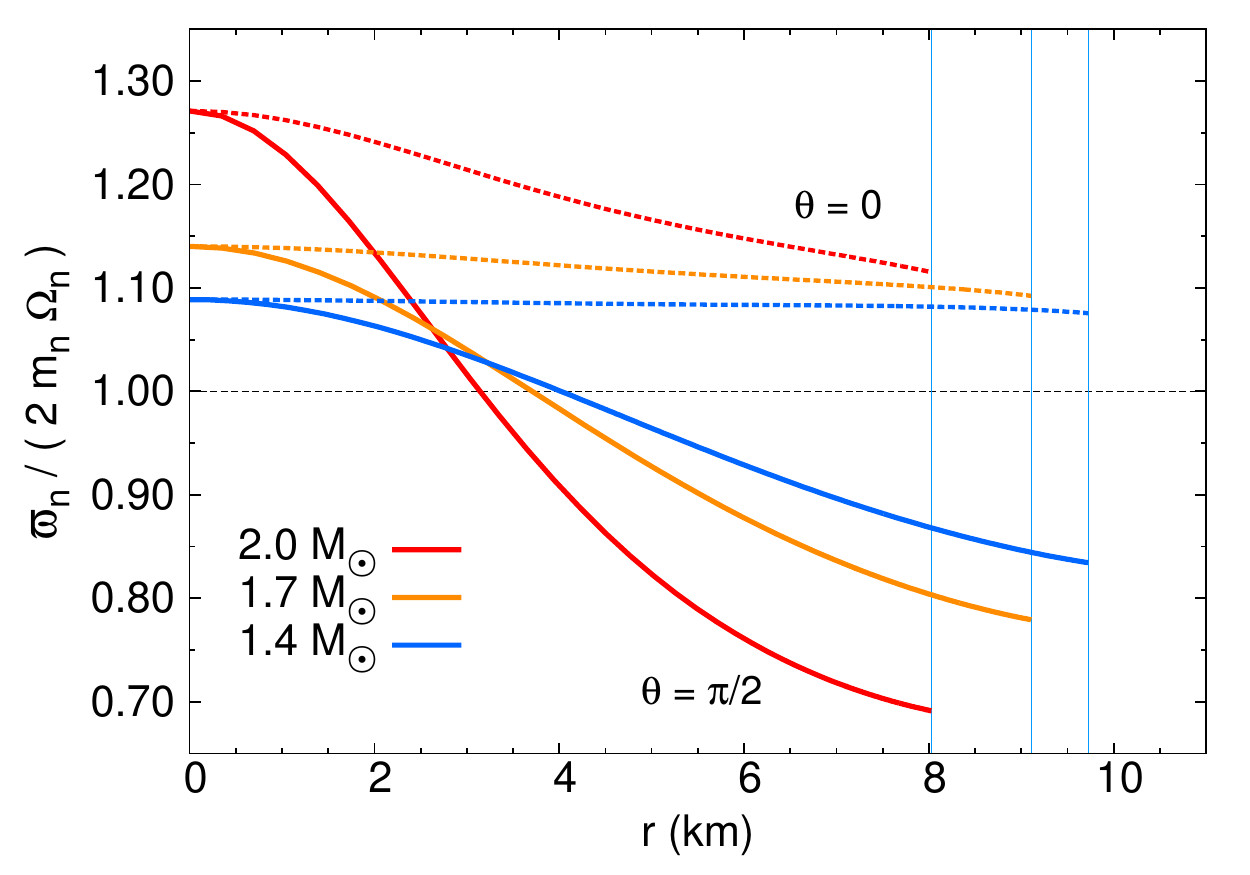}
   \caption{Superfluid vorticity in a neutron star plotted with respect to the radial
     coordinate $r$, in the equatorial ($\theta = \pi/2$) and polar
     ($\theta= 0$) planes, for $f_{\n} = f_{\p} =11.19$ Hz, assuming beta equilibrium. The neutron vorticity
     $\varpi_{\n}$ is normalized to its expected value
     $2\, m_{\n}\, \Omega_{\n}$ in the Newtonian limit, if corotation is enforced (black dashed lines). Blue vertical lines represent the radii at which $n_{\n}=0$, for both polar and equatorial planes given that the star is approximately spherical at such a low angular velocity.
     \textbf{Left:} Neutron vorticity profiles for a star
     with a gravitational mass $M_{\text{G}} = 1.4$ M$_{\odot}$, using DDH (blue lines) and DDH$\delta$ (orange lines) EoSs.  \textbf{Right:}  Influence of the gravitational mass of the star on the
     vorticity profiles, using the DDH EoS. Results in the equatorial (polar) plane are plotted in solid (dashed) lines in both panels. See the online version for colors.}
   \label{fig_vorticities}
 \end{figure*}

In Fig.~\ref{fig_vorticities}, the superfluid vorticity $\varpi_{\n}$ (\ref{def_vort}) is plotted as a function of the radial coordinate, for a star spinning at Vela's rotation frequency, \textit{i.e.} $f_{\n} = f_{\p}= 11.19$ Hz, and assuming beta equilibrium. The vorticity is normalized to its Newtonian limit $\varpi_{\n}^{\text{newt}}=2m_{\n}\Omega_{\n}$. In the left panel, the results from both EoSs are compared, whereas in the right panel vorticity profiles are represented for different gravitational masses, using the DDH EoS. 

The deviation from the Newtonian value can be simply interpreted in terms of the compactness parameter $\Xi$ of the star, defined as the dimensionless ratio of the gravitational mass $M_{\text{G}}$ of the star to its circumferential radius in the equatorial plane $R_{\text{c}, \text{eq}}$ (see \cite{gourgoulhon2010introduction} for definitions), \textit{i.e.}
\begin{equation}
\label{compactness}
\Xi = \frac{GM_{\text{G}}}{R_{\text{c}, \text{eq}}c^2} . 
\end{equation} 
For a 1.4 M$_{\odot}$ neutron star spinning at 11.19 Hz, the DDH$\delta$ EoS predicts a larger radius than the DDH one, which leads to $\Xi_{\text{DDH}} \gtrsim \Xi_{\text{DDH}\delta}$. The deviation from the non-relativistic case is therefore slightly stronger for DDH, as can be seen in the left panel of Fig.~\ref{fig_vorticities}. Similarly, since the compactness parameter increases with the
mass of the star, the deviation from the expected value in the
Newtonian limit is more important for more massive stars, and can reach a maximum of $\sim 30~\%$ (see the right panel of Fig.~\ref{fig_vorticities}). It is interesting to note that, since the compactness parameter is smaller for higher rotation rates at fixed gravitational mass, the general relativistic correction is found to be less important for stars spinning more rapidly.

Moreover, the quantity $\zeta$ involved in the spin-up time scale (\ref{tau_r}) can be determined from Eqs.~(\ref{def_kappa}) and (\ref{zeta}) using stationary configurations, by computing the superfluid vorticity profile and the appropriate moments of inertia. To reach high accuracy, the latter are calculated from Eq.~(\ref{def_I}) using a fourth-order finite difference method, either at given fluid baryon masses (case~i) or for a fixed total baryon mass with chemical equilibrium at the center (case~ii), depending on the assumption on the composition (see Sec.~\ref{chemical}). 

For a given rotation frequency, neglecting the lag and assuming beta equilibrium, $\zeta$ is found to decrease with increasing mass and tends towards $1$ for small compactness parameters (see Appendix~\ref{rise_newt}). In particular, for a star spinning at 11.19~Hz, $\zeta$ thus changes from 0.869 (0.880) for $M_{\text{G}} = 1.4$~M$_{\odot}$ to 0.763 (0.792) for a 2~M$_{\odot}$ neutron star with the DDH($\delta$) EoS. On the other hand, at fixed gravitational mass, $\zeta$ is approximately independent of the angular velocity in the slow-rotation approximation $\left(\Omega_{\n},\Omega_{\p} \ll \Omega_{\text{K}}\right)$. Since the moments of inertia are also nearly constant for low rotation rates,  $\tau_{\text{r}}$ turns out to be inversely proportional to the angular velocity, see Eq.~(\ref{tau_r}). However, for rotation frequencies substantially higher than that of Vela, the moments of inertia are found to increase more rapidly than $\kappa$, meaning that $\zeta$ decreases when the angular velocity gets 
higher. For instance, the DDH EoS leads to $\zeta= 0.834$ for a 1.4~M$_{\odot}$ neutron star spinning at 327~Hz, which is the rotation frequency of the fastest glitching pulsar observed so far (see the ATNF Pulsar Database\footnote{http://www.atnf.csiro.au/research/pulsar/psrcat.}; \citet{manchester2005australia}).

\subsection{Entrainment and frame-dragging effects}
\label{couplings}

\begin{figure*}
\includegraphics[width = 0.49\textwidth]{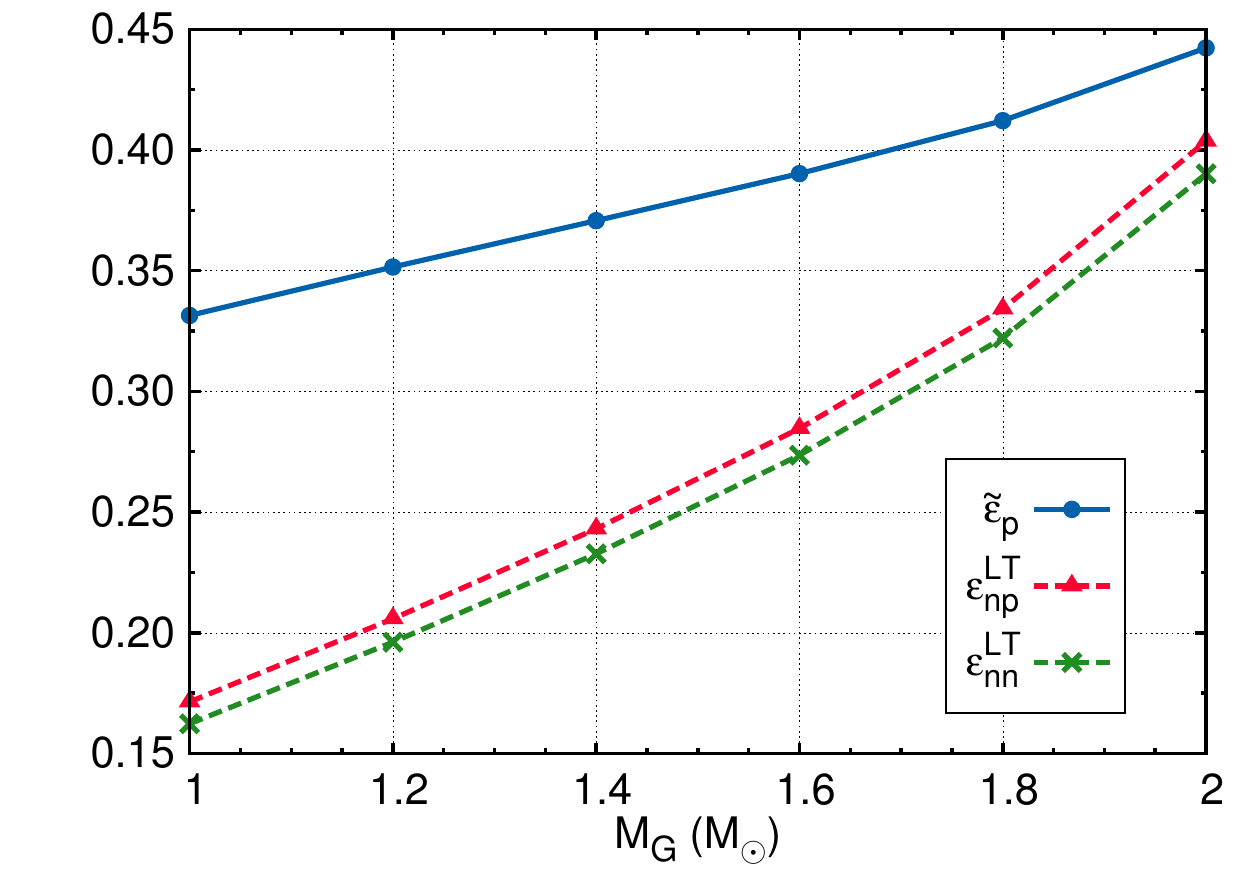}
\includegraphics[width = 0.49\textwidth]{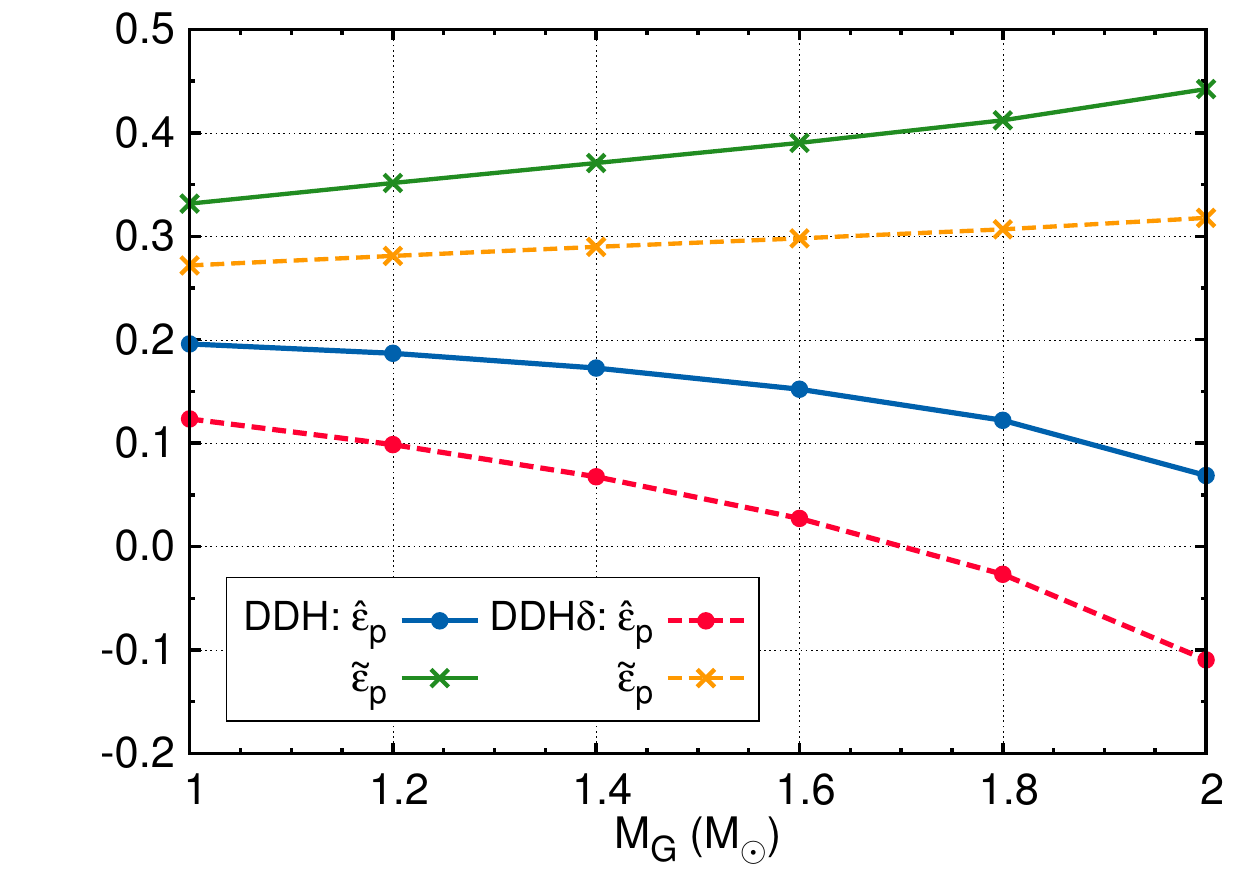}    
   \caption{\textbf{Left:} Entrainment parameter $\tilde{\varepsilon}_{\p}$ and Lense-Thirring coefficients $\varepsilon_{\n\!\p}^{\text{LT}}$ and $\varepsilon_{\n\!\n}^{\text{LT}}$ (see text and Appendix~\ref{mean_entr_par} for definitions) as functions of the gravitational mass $M_{\text{G}}$ of the star, for $f_{\n} = f_{\p}= 11.19$~Hz. Only results obtained from the DDH EoS are presented. \textbf{Right:} Total coupling coefficient $\hat{\varepsilon}_{\p}$ and entrainment parameter $\tilde{\varepsilon}_{\p}$ with respect to the gravitational mass, for a star spinning at 11.19~Hz, assuming corotation and beta equilibrium. Results obtained with the DDH($\delta$) EoS are shown in solid (dashed) lines.}
   \label{fig:Ipn}
 \end{figure*}

The cross moment of inertia $I_{\n \hspace*{-0.05 cm} \p}$ introduced in Eq.~(\ref{def_I}) contains all the possible couplings between neutrons and protons. A first coupling is due to \textit{entrainment}, which in the core of neutron stars comes from the strong interactions between nucleons. This effect was already discussed in details in Sec.~III of \cite{sourie2016numerical} for both DDH and DDH$\delta$ EoSs. In Newtonian gravity,  entrainment is the main fluid coupling at low angular velocities, see Appendix~\ref{I_newt}.  Nevertheless, as already mentioned in \citet{sourie2016numerical}, a new coupling arises in the general relativistic context from the so-called \textit{Lense-Thirring} or \textit{frame-dragging effect} (see \citet{carter1975application}).

In what follows, we characterize the total coupling between the fluids through the following quantities
 \begin{equation}
\label{eps_bar}
\hat{\varepsilon}_{\n} = \dfrac{I_{\n\hspace*{-0.05 cm}\p}}{\hat{I}_{\n}}  \ \ \ \text{and} \ \ \ \hat{\varepsilon}_{\p} = \dfrac{I_{\n\hspace*{-0.05 cm}\p}}{\hat{I}_{\p}}.
\end{equation}
From the stability conditions (\ref{condition}), these parameters are not arbitrary but must satisfy the following inequality
\begin{equation}
\left(\dfrac{1}{\hat{\varepsilon}_{\p}} - 1 \right) \left(\dfrac{1}{\hat{\varepsilon}_{\n}} - 1 \right) > 1.
\end{equation} 
At first order in the lag $\delta \Omega= \Omega_{\n} - \Omega_{\p}$ and in the slow-rotation approximation, these coupling coefficients can be written as 
\begin{equation}
\hat{\varepsilon}_X = \frac{\tilde{\varepsilon}_X- \varepsilon_{Y\!X}^{\text{LT}}}{1-\varepsilon_{Y\!X}^{\text{LT}}-\varepsilon_{X\!X}^{\text{LT}}},
\label{eps_bar_slow}
\end{equation}
where $X$ and $Y\neq X$ are n or p for neutrons or protons respectively, see Appendix~\ref{mean_entr_par} for details. In this expression, the term $\tilde{\varepsilon}_X$ characterizes entrainment effects averaged over the star (\ref{mean_entra_par_RG}), whereas $\varepsilon_{Y\!X}^{\text{LT}}$ and $\varepsilon_{X\!X}^{\text{LT}}$ represent respectively the frame-dragging effect on fluid $X$ caused by the second fluid and fluid $X$ itself, see  Eqs.~(\ref{mean_omega_RG}) and (\ref{LT_coupling_terms}). Quite remarkably, frame-dragging effects lead to similar fluid couplings as the entrainment arising from neutron-proton interactions, see Eq.~(\ref{J_RG}). Since the different quantities involved in Eq.~(\ref{eps_bar_slow}) are positive, the Lense-Thirring effect is found to act in an opposite way to entrainment in the core. The reason for the presence of minus signs in front of every quantity relative to frame-dragging effects is the following: a \textit{zero-angular-momentum observer} will rotate in the same sense as the whole star, leading a \textit{zero-angular-velocity observer} to have an angular momentum with an opposite sign to the total angular momentum 
of the star, even if this observer is static \citep{carter1975application}. As a consequence, in the absence of entrainment, the total coupling coefficient is still expected to be non-vanishing and negative. Although entrainment is likely to be small in the outermost regions of the core of neutron stars \citep{carter2006entrainment, chamel2006entrainment}, its overall effect on the whole star is not necessarily negligible and therefore $\hat{\varepsilon}_X$ could be positive or negative.

To assess the relative importance of these two effects to the total coupling coefficient $\hat{\varepsilon}_{\p}$, the quantities $\tilde{\varepsilon}_{\p}$ and $\varepsilon_{\n\!\p}^{\text{LT}}$ are plotted in the left panel of Fig.~\ref{fig:Ipn} as functions of the gravitational mass $M_{\text{G}}$ of the star for the DDH EoS.  Since typically $\delta \Omega \ll \Omega_{\n}, \Omega_{\p}$, we consider here that $\Omega_{\n}= \Omega_{\p}= 2\pi \times 11.19$~rad.s$^{-1}$. The mean proton entrainment parameter $\tilde{\varepsilon}_{\p}$ is increasing with the mass, because higher densities are reached in the star (see \cite{sourie2016numerical} - Fig.~2). Since general relativistic effects are the strongest for the most massive stars,  $\varepsilon_{\n\!\p}^{\text{LT}}$ increases with the stellar mass. It is interesting to note that $\tilde{\varepsilon}_{\p}$ and $\varepsilon_{\n\!\p}^{\text{LT}}$ are found to be roughly of the same order of magnitude, making the Lense-Thirring contribution to the total coupling be very 
important. A similar 
conclusion is reached with the DDH$\delta$ EoS.
Since $\varepsilon_{\n\!\n}^{\text{LT}}$ and $\varepsilon_{\n\!\p}^{\text{LT}}$ both characterize the contribution of the neutron fluid to frame-dragging effects, these two parameters are very close to each other, as can be seen in the left panel of Fig.~\ref{fig:Ipn}. For the same reasons, we have $\varepsilon_{\p\!\n}^{\text{LT}}\simeq \varepsilon_{\p\!\p}^{\text{LT}}$. Moreover, we numerically find that $\varepsilon_{\p\!\p}^{\text{LT}}~\simeq~\hat{I}_{\p}/\hat{I}_{\n} \times \varepsilon_{\n\!\n}^{\text{LT}}~\ll~\varepsilon_{\n\!\n}^{\text{LT}}$,  which means that the relative contribution of the two fluids to the frame-dragging effect is mainly due to their relative proportion in mass, as expected. The  typical values obtained for the  Lense-Thirring parameters are found to be consistent with the rough estimates given by \cite{carter1975application}, within a factor of $\sim 4$ for both EoSs.

 Using the DDH EoS, the equality of the cross moments of inertia $I_{\n\hspace*{-0.05 cm}\p}$ and $I_{\p\hspace*{-0.05 cm}\n}$, see Sec.~\ref{transfer}, is numerically verified with a precision better than $\sim 0.1\%$ for $f < 327$~Hz. In the following, we mainly focus on the proton coupling parameter $\hat{\varepsilon}_{\p}$ (\ref{eps_bar}). Indeed, the neutron parameter $\hat{\varepsilon}_{\n}$ can be simply deduced from Eq.~(\ref{eps_bar}), \textit{i.e.} $\hat{\varepsilon}_{\n}~=~\hat{I}_{\p} / \hat{I}_{\n} \times \hat{\varepsilon}_{\p} \ll\hat{\varepsilon}_{\p} $. For the numerical results displayed in Fig.~\ref{fig:Ipn}, we fixed the total baryon masses (case~ii). However, for the low rotation frequency we considered, fixing the individual baryon masses (case~i) would have lead to essentially the same results.

As displayed in the right panel of Fig.~\ref{fig:Ipn} for a star spinning at 11.19~Hz, the total coupling coefficient $\hat{\varepsilon}_{\p}$ is found to decrease significantly when the gravitational mass increases. This means that, although entrainment effects become more important, the frame-dragging contribution is increasing even more rapidly (see the left panel of Fig.~\ref{fig:Ipn}). The discrepancy between entrainment parameters and total coupling coefficients can be clearly seen in the right panel of Fig.~\ref{fig:Ipn}. For slowly rotating stars (with $f \lesssim 100$~Hz), results obtained from Eqs.~(\ref{eps_bar}) and (\ref{eps_bar_slow}) agree with a precision better than  $\sim 0.1\%$ for the DDH EoS and $\sim 0.3\%$ for the DDH$\delta$ one. As DDH and DDH$\delta$ predict stars with similar compactness parameters, frame-dragging effects are nearly the same for the two EoSs. Nevertheless, since entrainment effects are much stronger with DDH (see Fig.~2 of \citet{sourie2016numerical}), the total 
coupling is higher for DDH than for DDH$\delta$. 

Making use of the coupling parameters (\ref{eps_bar}), the general relativistic spin-up time scale (\ref{tau_r}) leads to a similar expression to that obtained within the Newtonian framework (see Eq.~(\ref{tau_r_newt})), namely
\begin{equation}
\label{tau_r_eps}
\tau_{\text{r}} = \dfrac{\hat{I}_{\p}}{\hat{I}}\times\frac{1 - \hat{\varepsilon}_{\p} -\hat{\varepsilon}_{\n}}{2\zeta\bar{\mathcal{B}} \Omega_{\n}}.
\end{equation}
It should be stressed however that general relativistic effects are not only included in the coefficient $\zeta$ but can also change the other parameters (for more details, see Sec.~\ref{GR}). Furthermore, using Eq.~(\ref{eps_bar_slow}) and considering that $\varepsilon_{\n\!\p}^{\text{LT}} \simeq \varepsilon_{\n\!\n}^{\text{LT}}$ and $\varepsilon_{\p\!\n}^{\text{LT}} \simeq \varepsilon_{\p\!\p}^{\text{LT}}$, the spin-up time scale is roughly given by 
\begin{equation}
\label{tau_r_slow}
\tau_{\text{r}} \simeq \frac{\hat{I}_{\p}}{\hat{I}}\times\frac{1 -\tilde{\varepsilon}_{\p} -\tilde{\varepsilon}_{\n}}{1 -\varepsilon_{\p\!\p}^{\text{LT}} -\varepsilon_{\n\!\n}^{\text{LT}}}\times \frac{1}{2\zeta\bar{\mathcal{B}} \Omega_{\n}}, 
\end{equation}
in the slow-rotation approximation. From this expression, we clearly see that the Lense-Thirring effect acts to slow down the angular momentum transfer, whereas entrainment effects in the core tend to make it more efficient. Still, note that
entrainment in the inner crust, which comes from Bragg scattering of dripped neutrons by nuclei \citep{chamel2005band, chamel2012neutron}, is expected to lead to $\tilde{\varepsilon}_{X}<0$. In such a case, both entrainment and frame-dragging effects shall contribute to slow down the glitch event.

\begin{figure}
\includegraphics[width = 0.49\textwidth]{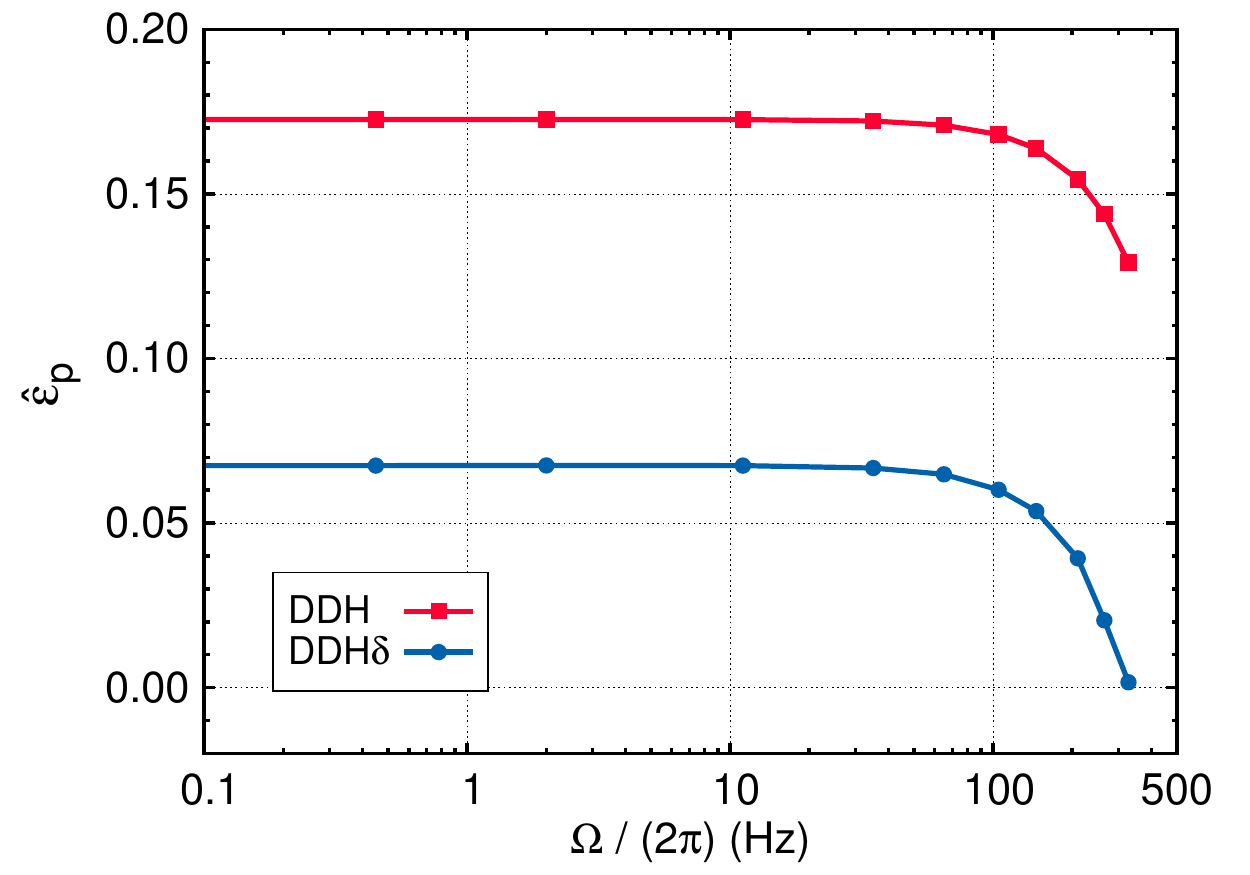}
   \caption{Coupling coefficient $\hat{\varepsilon}_{\p}$ computed from (\ref{eps_bar}) as a function of the rotation frequency for a 1.4~M$_{\odot}$ neutron star, assuming corotation and beta equilibrium (case ii). Results are shown in red (blue) for DDH($\delta$) EoS. The maximum rotation frequency considered here is 327~Hz, which corresponds to the highest frequency of observed glitching pulsars.}
   \label{fig:Inp_omega}
 \end{figure}

So far, we only focused on the slow-rotation approximation, in which the two fluids composing a neutron star are only coupled through entrainment and frame-dragging effects. Although this assumption is totally justified for Vela, it may not remain valid for other glitching pulsars. In particular, for rotation frequencies higher than  $\sim100$~Hz, the deformation of each fluid due to rotation leads to additional couplings via gravity, similarly to the Newtonian case (see Appendix~\ref{I_newt}). 
Beyond $\sim100$~Hz, the coupling coefficient $\hat{\varepsilon}_{\p}$ is found to decrease strongly when the rotation frequency increases, see Fig.~\ref{fig:Inp_omega}, and also depends on the assumption made on chemical equilibrium.

 \section{Numerical simulations of glitches}
 \label{numerical}

\subsection{Numerical procedure} 

\subsubsection{Computational scheme}
\label{discret}

Starting from two angular velocities $\Omega_{\n}^0$ and $\Omega_{\p}^0$ at the beginning of the glitch event (see Sec.~\ref{IC}), the evolution of the fluid rotation rates is computed from a series of equilibrium configurations (see Sec.~\ref{quasi_stat}), either keeping fixed the total baryon mass $M^{B}$ with $\mu^{\n}_c = \mu^{\p}_c$ or for constant partial baryon masses  $M^{B}_{\n}$ and $M^{B}_{\p}$, see Sec.~\ref{chemical}. This means that, for given angular velocities corresponding to the instant under consideration, the mutual friction torque on the right-hand side of Eq.~(\ref{evol_eq2}) is calculated at equilibrium, using the code described in \citet{sourie2016numerical}.  The moments of inertia involved in Eq.~(\ref{evol_eq2}) are computed either  through a fourth-order finite difference method or from a spectral interpolation based on Chebyshev polynomials. In the former, the moments of inertia are taken as constants during the glitch and evaluated at the rotation frequencies at the end of 
the glitch. On the contrary, in the latter, the moments of inertia are calculated for the angular velocities corresponding to the instant under consideration.

The angular velocities are evolved in time employing a
two-step explicit Adams-Bashforth method using a time step $\delta t\ll
\tau_{\text{r}}$. For the different results given in Sec.~\ref{results}, we typically consider time steps of the order of $\delta t\simeq \tau_{\text{r}}/10^{4} - \tau_{\text{r}}/10^{2}$, where $\tau_{\text{r}}$ is estimated from Eq.~(\ref{tau_r}). Note that, for the different simulations performed, the total baryon mass  $M^{B}$ and the total angular momentum $J$ are conserved during the glitch with a precision better than~$10^{-10}$.

\subsubsection{Initial conditions}
\label{IC}

The lag $\delta \Omega_0 = \Omega_{\n}^0 - \Omega_{\p}^0 $ at the beginning of the glitch should be determined from the pinning of vortex lines during the pre-glitch evolution (see, e.g., \cite{haskell2012modelling}). Still, it is possible to get the initial angular velocities $\Omega_{\n}^0$ and $\Omega_{\p}^0$ from basic considerations, regardless of the physical processes that build up the lag and trigger
the glitch, as described in the following.

 \begin{figure*}
\includegraphics[width = 0.49\textwidth]{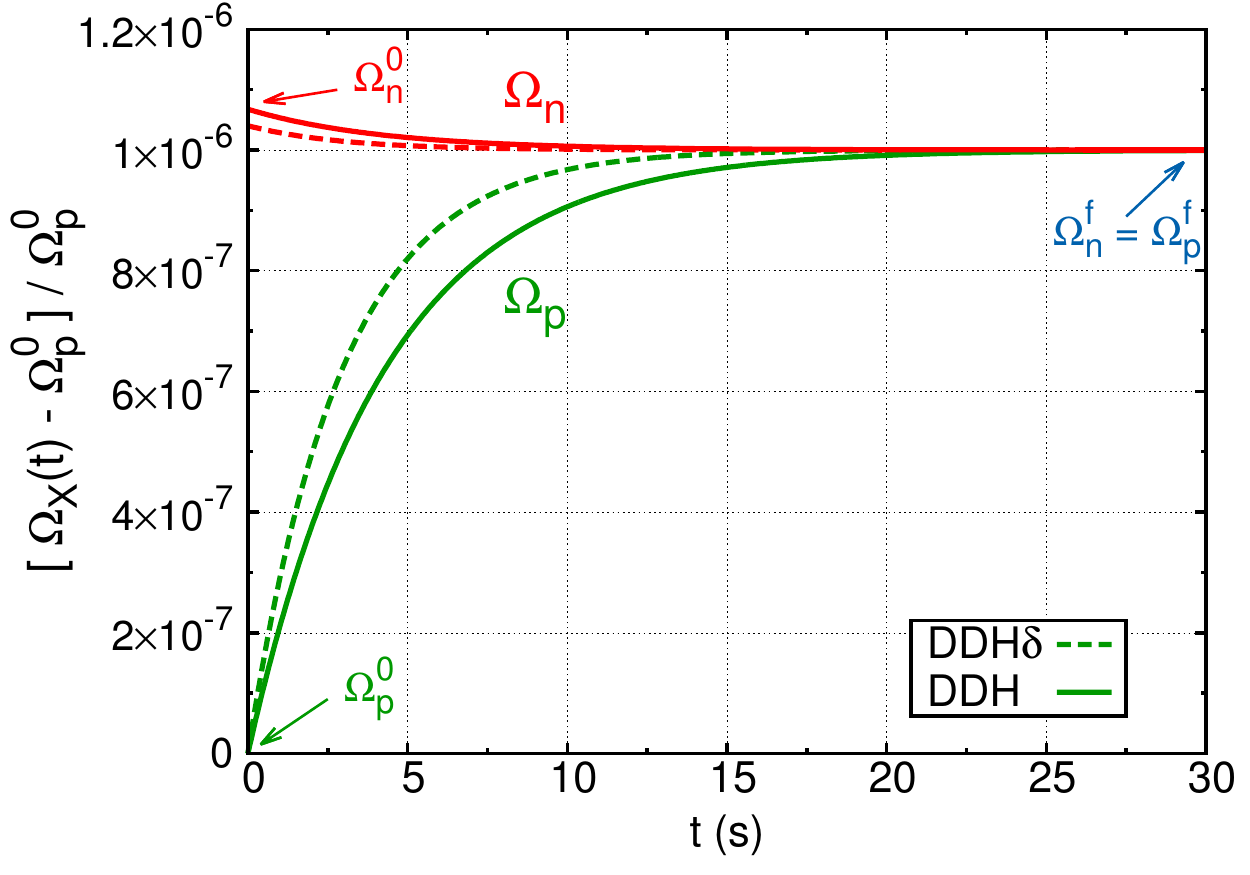}
\includegraphics[width = 0.49\textwidth]{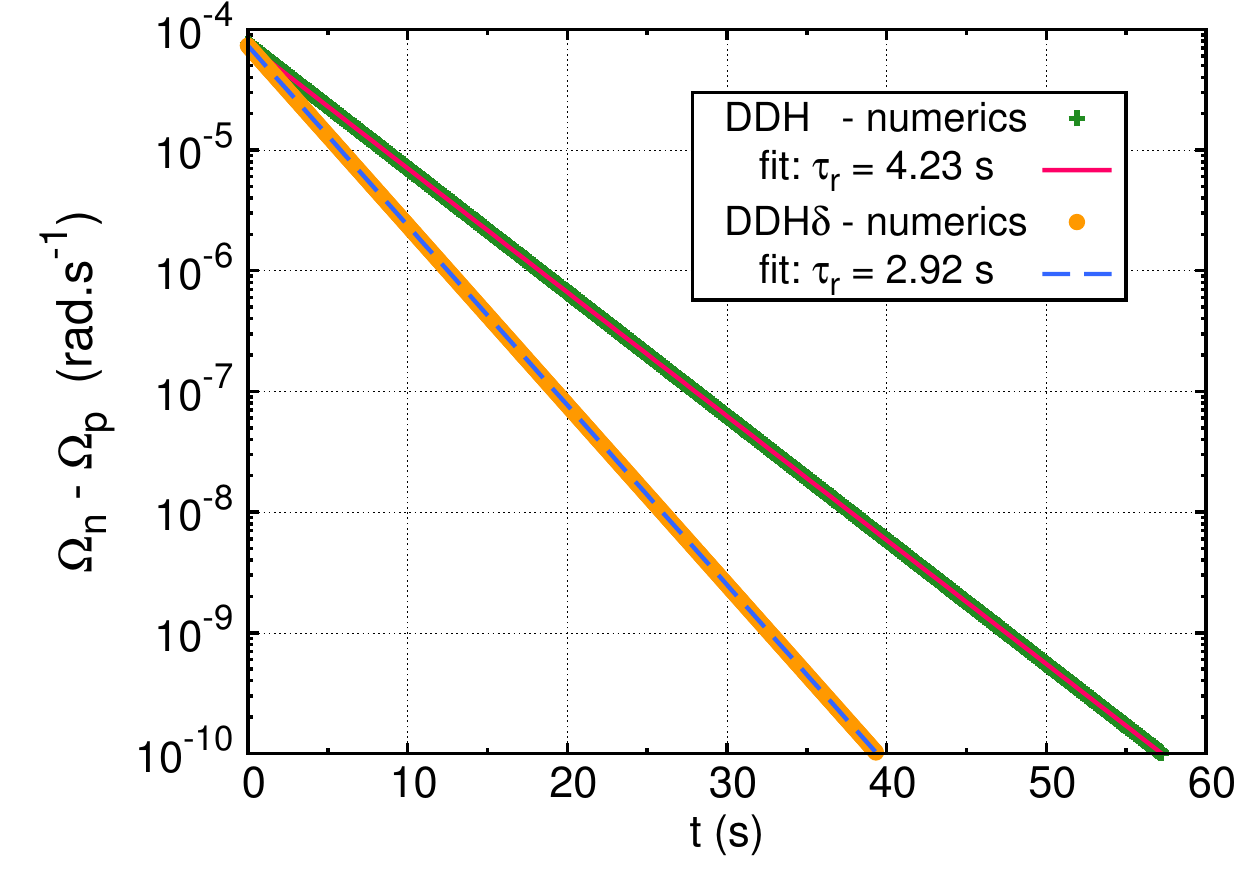}
   \caption{\textbf{Left:} Time evolution of the superfluid and proton angular velocities for realistic input parameters corresponding to Vela $\left(f= 11.19~\text{Hz, } \Delta\Omega/\Omega=10^{-6}\right)$. The gravitational mass is $M_\text{G} = $ 1.4 M$_{\odot}$. The mutual friction parameter is $\bar{\mathcal{B}}=10^{-4}$. The dynamical evolution of the two fluids is plotted in terms of $(\Omega_X(t) -\Omega^0_{\p})/\Omega^0_{\p}$. Results obtained from the DDH($\delta$) EoS are shown in solid (dashed) lines.  \textbf{Right:} Lag  $\delta\Omega= \Omega_{\n} - \Omega_{\p}$  as a function of time for the same input parameters. Straight lines correspond to fits with exponential decaying laws.}
   \label{fig:evol_lag}
 \end{figure*}

We denote by $\Omega_{\n}^{\text{f}}$ and $\Omega_{\p}^{\text{f}}$ the angular velocities at the end of the spin-up, \textit{i.e.} when
the post-glitch relaxation starts. By assuming that the system relaxes completely during the glitch rise, as suggested by Eq.~(\ref{lag_evol}), we have
\begin{equation}
\Omega_{\n}^{\text{f}} =\Omega_{\p}^{\text{f}} = \Omega^{\text{f}}.
\end{equation}
The initial proton rotation frequency $\Omega_{\p}^0$  can be simply deduced from a given choice of the amplitude
\begin{equation}
\frac{\Delta \Omega}{\Omega}=\frac{\Omega^{\text{f}} - \Omega_{\p}^0}{\Omega_{\p}^0}
\end{equation} 
of the glitch to be modelled, which is typically of the order of  $\sim 10^{-6}$ for Vela  \citep{dodson2007two}. 
Furthermore, the total angular momentum $J = J_{\n} + J_{\p}$ being
conserved during the whole glitch event, see Eq.~(\ref{evol_eq}), the last
unknown $\Omega_{\n}^0$ can be determined from
\begin{equation}
J\left( \Omega_{\n}^0, \Omega_{\p}^0\right) = J\left( \Omega_{\n}^{\text{f}}, \Omega_{\p}^{\text{f}}\right).
\end{equation}
Since the glitch amplitudes are extremely small, the final angular velocities
$\Omega_{X}^{\text{f}}$ are very close to the initial values
$\Omega_{X}^0$.  Expanding $J$ to first order in the angular velocities is thus sufficient to determine $\Omega_{\n}^0$ with very high accuracy:
 \begin{equation}
 \Omega_{\n}^{0} \simeq\Omega^{\text{f}}\left( 1+ \dfrac{\hat{I}_{\p} }{\hat{I}_{\n}} \frac{\Delta \Omega}{\Omega} \right),
 \end{equation}
where the moments of inertia, defined by (\ref{def_I}), are computed at the end of the glitch, \textit{i.e.} for $\Omega_{\p}^{\text{f}} = \Omega_{\n}^{\text{f}} $. 

The initial lag $\delta \Omega_0$ is therefore given by
\begin{equation}
\label{lag_init}
\delta \Omega_0 = \Omega_{\n}^{0} - \Omega_{\p}^{0} \simeq \Omega^{\text{f}} \dfrac{\hat{I}}{\hat{I}_{\n} } \frac{\Delta \Omega}{\Omega},
\end{equation}
which leads to $\sim 7.10^{-5}$ rad.s$^{-1}$ for the Vela pulsar, taking $\Omega^{\text{f}}/(2\pi) = 11.19$ Hz. We deduce that the lag between the two fluids always verifies the condition
\begin{equation}
\label{corot}
\frac{\delta \Omega}{\Omega} \leqslant \frac{\delta \Omega_0}{\Omega} \simeq \frac{\Delta \Omega}{\Omega} \ll 1,
\end{equation}
meaning that the deviation from corotation remains very small during the glitch event.

To summarize, the numerical simulations require the following macroscopic ingredients:
\begin{itemize}
\item the rotation rate $\Omega^{\text{f}}$ of the star,
\item its gravitational mass $M_{\text{G}}$,
\item the glitch amplitude $\Delta \Omega / \Omega$,  
\end{itemize}
which can be potentially directly obtained from observations. In addition, the following microscopic inputs need to be specified:
\begin{itemize}
\item the EoS used to describe the interior of the star (for the adopted composition as discussed in Sec.~\ref{chemical}),
\item the mutual friction parameter $\bar{\mathcal{B}}$.
\end{itemize}   

Contrary to the total baryon mass $M^{B}$, the gravitational mass $M_{\text{G}}$, which corresponds to the observed mass of the pulsar, should vary during the glitch spin up. Nevertheless, we note that the change in $M_{\text{G}}$ associated with the angular momentum transfer is found to be smaller than a few $10^{-10}$ for the different tests performed with $\Delta \Omega / \Omega = 10^{-6}$. 

 \subsection{Numerical results}
 \label{results}
 
We now present various results obtained from the numerical simulations described in the previous section. In the following, we mainly consider slowly rotating pulsars like Vela for which the assumption on the composition is unimportant (for the actual numerical calculations, we consider case~ii, see Sec.~\ref{chemical}). The impact of the chemical equilibrium on the evolution of more rapidly rotating neutron stars is discussed at the end of Sec.~\ref{time}.

\subsubsection{Dynamical evolution}
\label{time}

In Fig.~\ref{fig:evol_lag}, we show the temporal evolution
of the two angular velocities and the lag $\delta\Omega = \Omega_{\n} -
\Omega_{\p}$ for conditions corresponding to the Vela pulsar. The gravitational mass of this pulsar being not (well) known, we have chosen the 
canonical value of $M_\text{G} = 1.4 $ M$_\odot$ in this example. Moreover, the mutual friction parameter is arbitrarily fixed to $\bar{\mathcal{B}}=10^{-4}$. 
Results are shown in Fig.~\ref{fig:evol_lag} for both EoSs studied in \citet{sourie2016numerical}, namely the DDH and DDH$\delta$ EoSs, with only small differences between both. As can be seen in the right panel, the evolution of the lag can be very well described by an exponential law of the form (\ref{lag_evol}), as expected from Sec.~\ref{estimate}. For the present example, we find $\tau_r = 4.23$~s with the DDH EoS and $\tau_r = 2.92$~s with the DDH$\delta$ one. It should be remarked that these characteristic times, obtained from the time evolution of the lag, correspond indeed to the spin-up time scales that could be measured from precise timing observations of glitches, see Eq.~(\ref{time_evol}). 

    \begin{figure}
\includegraphics[width = 0.49\textwidth]{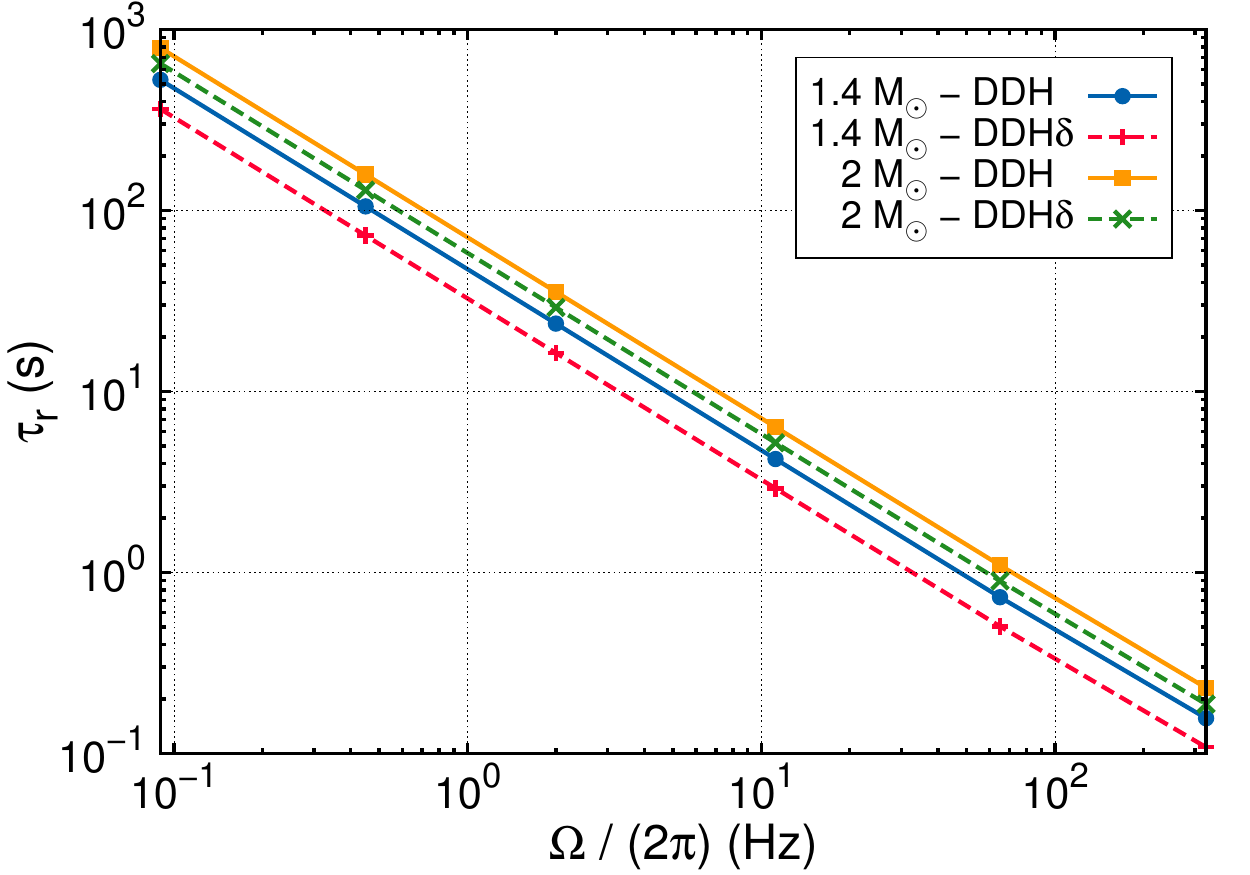}
   \caption{Glitch rise time $\tau_{\text{r}}$ as a function of the pulsar rotation rate $f=\Omega/(2\pi)$, for a glitch amplitude $\Delta \Omega / \Omega = 10^{-6}$. Results obtained from the DDH($\delta$) EoS are shown in solid (dashed) lines. Configurations with two different gravitational masses are displayed, using $\bar{\mathcal{B}}= 10^{-4}$.}
   \label{fig:tau_r_omega}
 \end{figure}

In order to study the dependence of the rise time on the different  input parameters and to compare with the results given in  Sec.~\ref{estimate}, we have performed a series of simulations, varying in particular the mutual friction parameter $\bar{\mathcal{B}}$, the rotation rate $\Omega$ of the star as well as its gravitational mass $M_{\text{G}}$. In Fig.~\ref{fig:tau_r_omega}, the spin-up time scale is plotted with respect to the pulsar angular velocity $\Omega^{\text{f}}$, for two different gravitational masses and both EoSs, assuming  a glitch amplitude $\Delta \Omega / \Omega = 10 ^{-6}$. The rise time turns out to be inversely proportional to the pulsar rotation rate with a high accuracy, consistently with (\ref{tau_r}). A very small deviation from this simple behaviour can be seen for $f \gtrsim 100$~Hz, due to the strong increase of the moments of inertia with the angular velocity in this range of values (see Sec.~\ref{sup_vorti}).

For the different cases considered in Fig.~\ref{fig:tau_r_omega}, the numerical results are found to agree with values inferred from Eq.~(\ref{tau_r}) with a precision better than $\sim 5\times 10^{-6}$. This limit of accuracy comes from the numerical errors  associated with dynamical simulations, which are dominated by the discretization in time and the precision with which the moments of inertia are computed. The reason why Eq.~(\ref{tau_r}) gives  such a precise estimate for the spin-up time scale comes from the extremely small glitch amplitudes that are observed. The spin-up time scale can thus be very precisely estimated from Eq.~(\ref{tau_r}) by merely computing stationary configurations and ignoring the change in the moments of inertia during the glitch.

 \begin{figure}
\includegraphics[width = 0.49\textwidth]{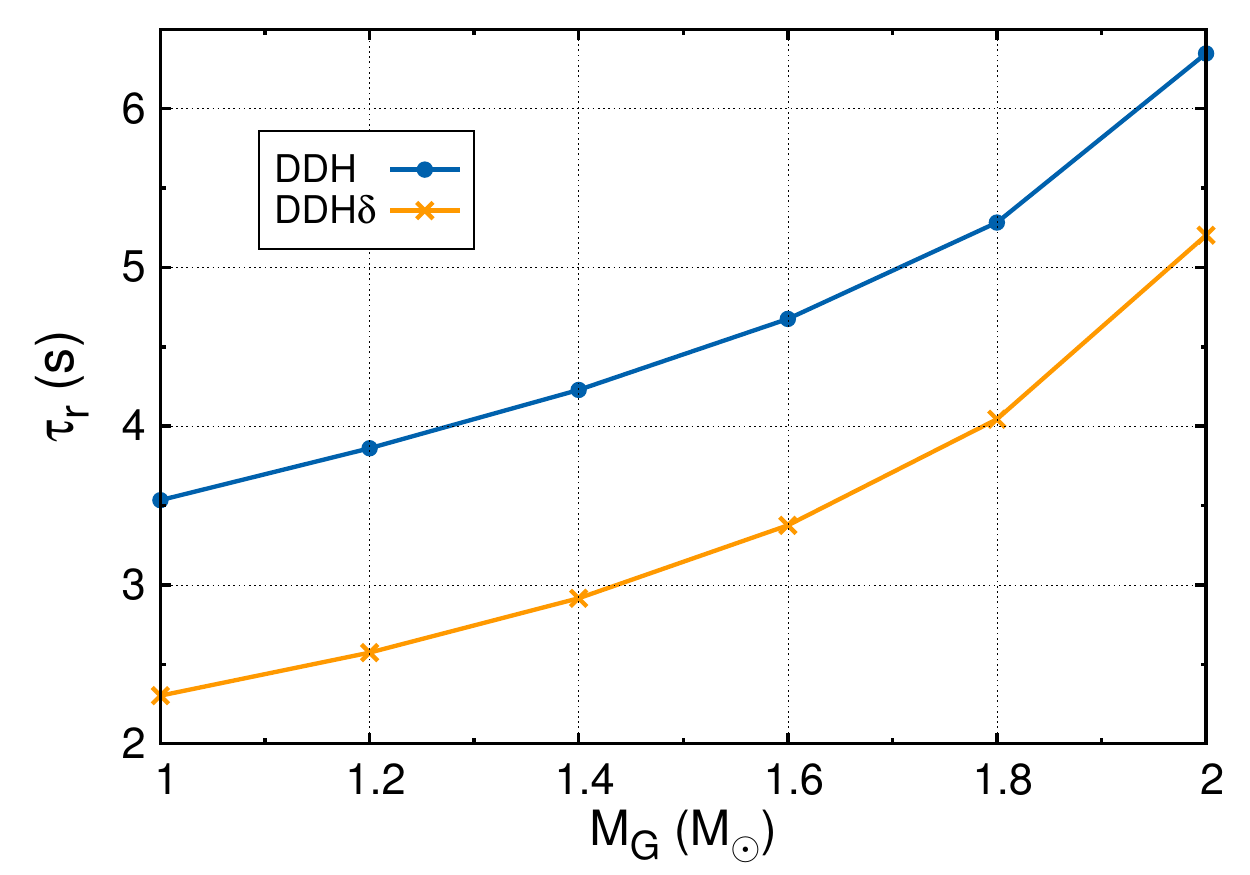}
   \caption{Glitch rise time $\tau_{\text{r}}$ plotted with respect to the gravitational mass $M_{\text{G}}$, for a star spinning at 11.19 Hz. The glitch parameters $\bar{\mathcal{B}}$ and $\Delta \Omega / \Omega$ are respectively fixed to $10^{-4}$ and $10^{-6}$. Results obtained from the DDH($\delta$) EoS are displayed with blue dots (yellow crosses). }
   \label{fig:tau_r_mass}
 \end{figure}

All other parameters fixed, the rise time increases with the gravitational mass of the star, as highlighted in Fig.~\ref{fig:tau_r_mass}. Several reasons can be invoked to explain this fact. First, the proton fraction $x_{\p}=n_{\p}/\left(n_{\n}  + n_{\p}\right)$ and therefore the ratio $\hat{I}_{\p} /\hat{I}$ are strongly increasing with the mass of the star, see Appendix~\ref{xp}.  As the neutron fraction decreases, the transfer of angular momentum becomes longer (see Eq.~(\ref{tau_r_eps})). Moreover, the coupling coefficient $\hat{\varepsilon}_{\p}$ and the quantity $\zeta$ are also found to decrease significantly as the gravitational mass increases, see Sec.~\ref{stat_conf}. The transfer of angular momentum is thus slowed down (Eq.~(\ref{tau_r_eps})). It should also be noticed that, even if the coupling is much stronger for DDH (see the right panel of Fig.~\ref{fig:Ipn}), the spin-up time scale is  systematically longer with the DDH EoS than with the DDH$\delta$ one because the proton fraction and thus  
the ratio $\hat{I}_{\p} /\hat{I}$ predicted by this EoS are much higher.

Finally, a few tests have been also performed to study the influence of the assumption  concerning the evolution of the chemical composition during the glitch rise (see Sec.~\ref{chemical}). Whether considering constant individual baryon masses (case~i) or fixed total baryon mass with beta equilibrium at the center (case~ii) leads to  negligible differences for the glitch rise time at low rotation frequencies. For instance, the deviation is lower than $\sim 7\times10^{-5}$ for a star rotating at 11.19~Hz within both EoSs, assuming a small glitch amplitude and fixed moments of inertia. Nevertheless, the impact of the assumption on the chemical equilibrium increases sharply with the angular velocity: for both EoSs, the discrepancy between cases (i) and (ii) is of the order of $\sim 2\times10^{-3}$ for 65 Hz and $\sim 5\times10^{-2}$ for 327~Hz.  Still, it is important to note that the influence of the assumption relative to chemical equilibrium is much smaller than the dependence of the rise time on any other input 
parameters of our model, such as masses or rotation rates. This is the reason why we consider only case (ii) in most of the results presented in the present paper.

 \subsubsection{Contribution of general relativity}
\label{GR}

\begin{table}
	\centering
  \caption{Polytropic parameters defining EoSs I (\ref{EoSI}) and II (\ref{EoSII}). For EoS~I, $\kappa_{\n}$, $\kappa_{\p}$ and $\kappa_{\n\hspace*{-0.05 cm} \p}$ are given in units of $\rho_{0}c^2n_0^{-2}$ where $n_0 = 0.1$ fm$^{-3}$ and $\rho_{0}= 1.66\times10^{17}$ kg.m$^{-3}$, whereas $\kappa_{\n}$ and $\kappa_{\p}$ are respectively expressed in units of $\rho_{0}c^2n_0^{-2.1}$ and $\rho_{0}c^2n_0^{-2.3}$ for the second EoS. For both EoSs, $\kappa_{\Delta}$ is expressed in units of $\rho_0n_0^{-2}$.}
  \label{tab:EoS}
\begin{tabular}{c|cccc}\hline
& $\kappa_{\n}$ & $\kappa_{\p}$ & $\kappa_{\n\hspace*{-0.05 cm} \p}$  & $\kappa_{\Delta}$   \\[2 pt]    \hline \\[-5 pt] 
     EoS I & 0.05 & 0.5 &  0.025 & 0.02   \\[2 pt]
      EoS II & 0.046 & 1.4 & - &0.1 \\
      \hline
\end{tabular}
\end{table}

    \begin{figure*}
\includegraphics[width = 0.49\textwidth]{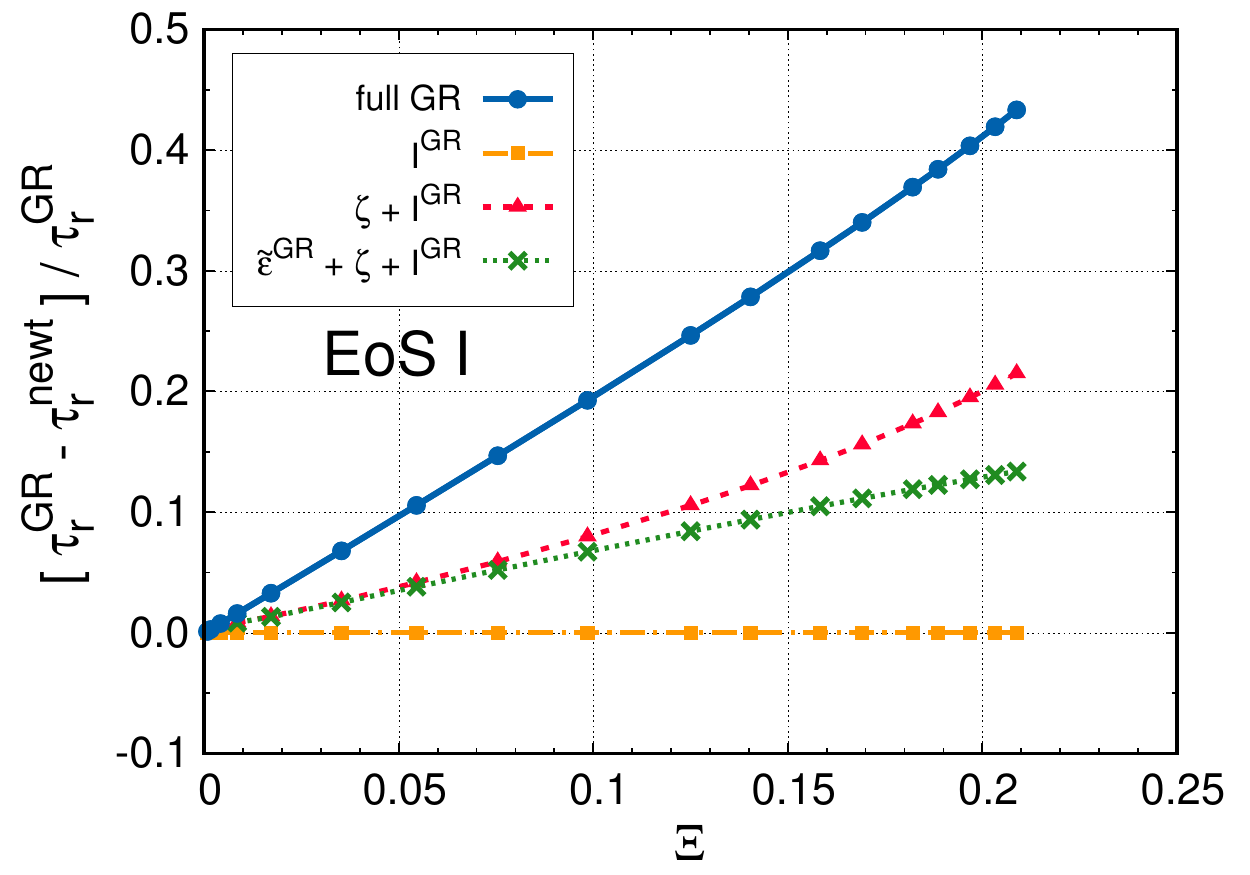}
\includegraphics[width = 0.49\textwidth]{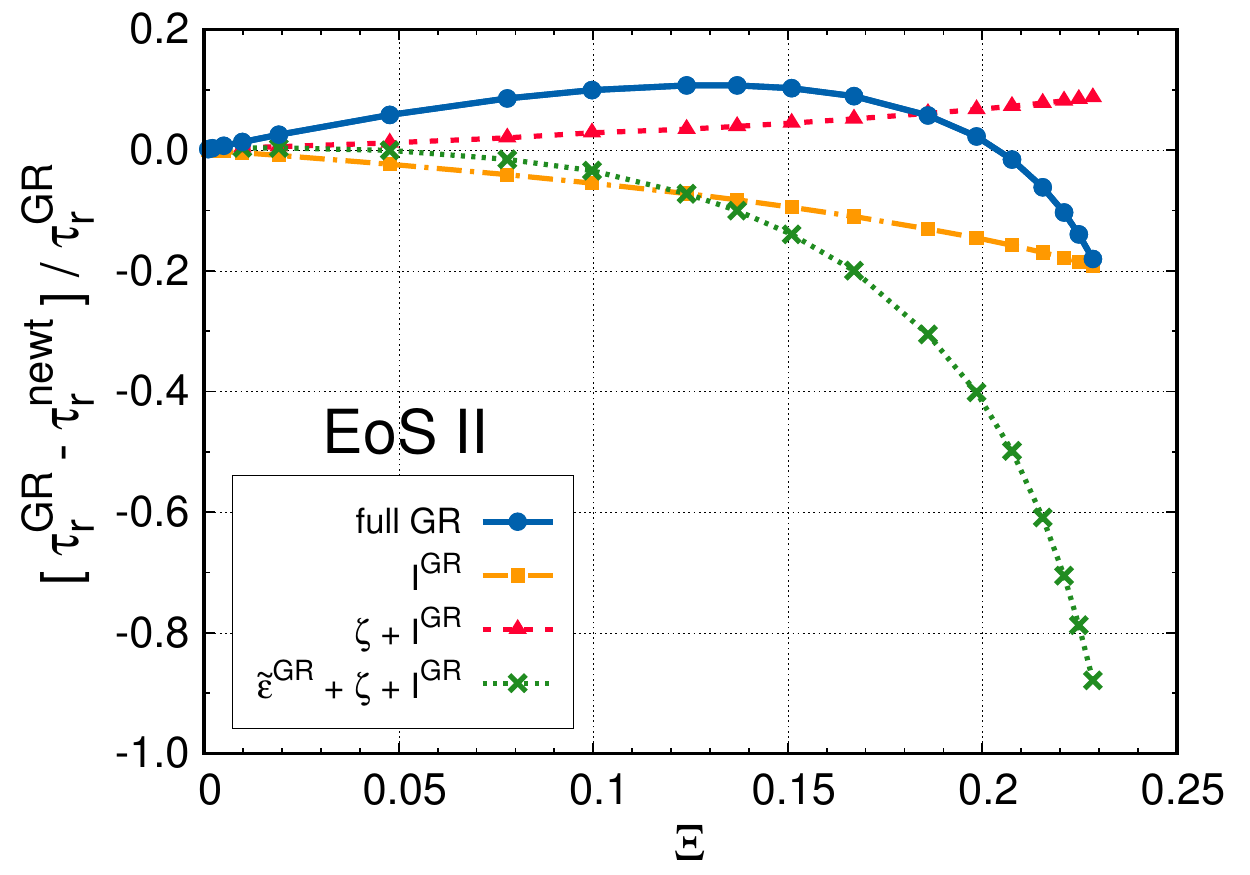}
   \caption{Relative differences between general relativistic and Newtonian rise times (blue solid lines) for a glitch amplitude $\Delta \Omega / \Omega = 10^{-6}$ as functions of the (relativistic) compactness parameter, for a star spinning at 10~Hz. Results are shown for the two polytropic EoSs described in the text: the left (right) panel corresponds to EoS~I (EoS~II). Other curves highlight the contribution of general relativity in the computation of the different terms involved in the spin-up time scale~(\ref{tau_r_eps}): results accounting only for the general relativistic corrections on the ratio $\hat{I}_{\p} /\hat{I}$ are displayed in orange dashed-dotted lines, whereas general relativistic corrections on both $\hat{I}_{\p} /\hat{I}$ and $\zeta$ are considered in the red dashed lines. The green dotted lines include the additional corrections on the entrainment parameters $\tilde{\varepsilon}_X$ but do not account for the frame-dragging contribution to the coupling coefficients $\hat{\varepsilon}_X$ (\ref{eps_bar_slow}). 
     }
   \label{fig:comp}
 \end{figure*}

To study the global contribution of general relativity to the spin-up time scale, we compare the rise times obtained within both relativistic and Newtonian frameworks. For simplicity, we consider polytropic EoSs, as implemented by \cite{prix2005relativistic}.

In Fig.~\ref{fig:comp}, the relative differences on $\tau_{\text{r}}$ are plotted with respect to the compactness parameter (\ref{compactness}), obtained for a star rotating at 10 Hz by varying its (gravitational) mass. We consider two different EoSs (referred to as EoS~I and II), with small and respectively strong entrainment effects, associated with the following Lagrangian densities $\Lambda=-\mathcal{E}$ with
 \begin{equation}
\mathcal{E} =  \frac{1}{2}\kappa_{\n}n_{\n}^2 + \frac{1}{2}\kappa_{\p}n_{\p}^2+ \kappa_{\n\hspace*{-0.05 cm} \p}n_{\n}n_{\p}+ \kappa_{\Delta}n_{\n}n_{\p}\Delta^2 \ \left(\ +\ \rho c^2\ \right), 
\label{EoSI}
\end{equation}
and 
 \begin{equation}
\mathcal{E} =  \frac{1}{2}\kappa_{\n}n_{\n}^{2.1} + \frac{1}{2}\kappa_{\p}n_{\p}^{2.3}+ \kappa_{\Delta}n_{\n}n_{\p}\Delta^2 \ \left(\ +\ \rho c^2\ \right), 
\label{EoSII}
\end{equation}
respectively, where $\rho = m_{\n}n_{\n} + m_{\p} n_{\p}$, $\Delta$ stands for the relative speed between the fluids and we have used the same notation as in \cite{prix2005relativistic}. Note that the rest mass energy density  is only present in the general relativistic case (more details on the differences between Newtonian and relativistic computations can be found in Sec.~IV-A of \cite{prix2005relativistic}). The different parameters $\kappa_{\n}$, $\kappa_{\p}$ and $\kappa_{\n\hspace*{-0.05 cm} \p}$ (see Table~\ref{tab:EoS}) are chosen in order to reproduce ``realistic" values for the mass, radius and proton fraction $x_{\p}$ of the stars: for instance, EoS~I leads to a constant proton fraction throughout the star with $x_{\p}= 0.05$, whereas a varying proton fraction in the range $x_{\p}\simeq 0.05 - 0.1$  is obtained with EoS~II for a 1.4 M$_{\odot}$ (relativistic) neutron star spinning at 10~Hz. Both EoSs predict an external circumferential radius in the equatorial plane $R_{\text{c, \!eq}}\simeq 13$ km for a relativistic neutron star with $M_{\text{G}}=1.4$~M$_{\odot}$. 
The entrainment contribution is included through the coefficient $\kappa_{\Delta}$. The forms of the EoSs are taken consistently with the fact that entrainment effects should vanish when one of the fluids disappears (see \citet{sourie2016numerical}). For EoS I, we take $\kappa_{\Delta}=0.02$
in order to satisfy all the required stability conditions  \citep{chamel2006entrainment}. This choice leads to $\tilde{\varepsilon}_{\p}\simeq0.07$ for a (relativistic) neutron star spinning at 10 Hz, with $M_{\text{G}}$ = 1.4~M$_{\odot}$. Although this value is quite small compared to realistic EoSs (Fig.~\ref{fig:Ipn}), it still corresponds to the outer core of neutron stars, where entrainment effects 
are nearly vanishing. On the contrary, requiring stability for EoS~II, a value of $\kappa_{\Delta}=0.1$ leads to $\tilde{\varepsilon}_{\p}\simeq0.29$ for the same mass and spin, which happens to be much more realistic.

General relativity is  expected to play a role in determining the moments of inertia $\hat{I}_{\p}$ and $\hat{I}$, the quantity $\zeta$ and the coupling coefficients $\hat{\varepsilon}_X$ involved in the spin-up time scale (\ref{tau_r_eps}). To highlight the contribution of general relativity on these different terms, the following quantities are plotted in Fig.~\ref{fig:comp} for EoS~I (left panel) and EoS~II (right panel):
\begin{itemize}
\item $\dfrac{\tau_{\text{r}}^{\text{GR}} -\tau_{\text{r}}^{\text{newt}} }{ \tau_{\text{r}}^{\text{GR}}} =1 - \dfrac{\hat{I}_{\p}^{\text{newt}}}{\hat{I}^{\text{newt}}}\dfrac{\hat{I}^{\text{GR}}}{\hat{I}_{\p}^{\text{GR}}}\dfrac{1-\tilde{\varepsilon}_{\p}^{\text{ newt}} - \tilde{\varepsilon}_{\n}^{\text{ newt}}}{1-\hat{\varepsilon}_{\p}^{\text{ GR}} - \hat{\varepsilon}_{\n}^{\text{ GR}}}\zeta $, denoting Newtonian and general relativistic quantities by ``newt" and ``GR" respectively (see Eq.~(\ref{eps_bar_slow}) and Appendix~\ref{mean_entr_par} for the definitions of the total coupling coefficients $\hat{\varepsilon}_X$ and the entrainment parameters $\tilde{\varepsilon}_X$),
\vspace*{0.2 cm}
\item $ 1 - \dfrac{\hat{I}_{\p}^{\text{newt}}}{\hat{I}^{\text{newt}}}\dfrac{\hat{I}^{\text{GR}}}{\hat{I}_{\p}^{\text{GR}}}$, accounting only for the relativistic corrections on the ratio $\hat{I}_{\p}/\hat{I}$,
\vspace*{0.2 cm}
\item $1 - \dfrac{\hat{I}_{\p}^{\text{newt}}}{\hat{I}^{\text{newt}}}\dfrac{\hat{I}^{\text{GR}}}{\hat{I}_{\p}^{\text{GR}}}\zeta $, considering both relativistic corrections on the moments of inertia and $\zeta$,
\vspace*{0.2 cm}
\item  $1 - \dfrac{\hat{I}_{\p}^{\text{newt}}}{\hat{I}^{\text{newt}}}\dfrac{\hat{I}^{\text{GR}}}{\hat{I}_{\p}^{\text{GR}}}\dfrac{1-\tilde{\varepsilon}_{\p}^{\text{ newt}} - \tilde{\varepsilon}_{\n}^{\text{ newt}}}{1-\tilde{\varepsilon}_{\p}^{\text{ GR}}-\tilde{\varepsilon}_{\n}^{\text{ GR}}}\zeta $, taking into account relativistic corrections on the  moments of inertia, $\zeta$ and the entrainment parameters $\tilde{\varepsilon}_{X}$.
\end{itemize}

As expected, general relativistic corrections tend to zero, when the compactness parameter decreases. Concerning EoS~I, general relativistic corrections on the ratio  $\hat{I}_{\p} / \hat{I}$ are found to be extremely small (see the left panel of Fig.~\ref{fig:comp}). This is due to the fact that $x_{\p}$ is constant throughout the whole star within this EoS and $\hat{I}_{\p} / \hat{I}=x_{\p}$ in both Newtonian and general relativistic frameworks, for a slowly rotating star in beta equilibrium and in the limit of vanishing lag between the fluids. In Newtonian gravity, $\zeta$ is equal to~1. Since general relativity leads to $\zeta\lesssim 1$ (see Sec.~\ref{sup_vorti}), this quantity acts to lengthen the rise time. Moreover, the general relativistic entrainment parameters are found to be much higher than their Newtonian counterparts, because higher densities are reached when general relativity is considered. These general relativistic corrections on entrainment tend to lower the rise time. Finally, frame-dragging contribution to the coupling coefficients $\hat{\varepsilon}_{X}$  (\ref{eps_bar_slow}) also leads to a longer spin-up time scale (see Eq.~(\ref{tau_r_slow})). Similar remarks apply to EoS~II but with two differences. First, general relativity slightly modifies the ratio of the moments of inertia, see the right panel of Fig.~\ref{fig:comp}. Furthermore, entrainment effects are much more important within this EoS than in EoS~I, leading to a much larger reduction of the rise time. To summarize, general relativistic corrections on the different terms involved in the spin-up time scale are  found to be roughly  of the same order of magnitude but depend strongly on the EoS considered. In particular, the frame-dragging contribution to the fluid couplings is found to be important.


For values of the compactness parameter relevant for neutron stars, \textit{i.e.} $\Xi \sim 0.15- 0.20$, these two EoSs predict that an error of the order of $\sim 20 - 40\%$ is made on the rise time by using Newtonian gravity instead of general relativity, as can be seen in Fig.~\ref{fig:comp}. It is therefore necessary to account for general relativistic effects in order to get precise results on the spin-up time scales. Furthermore, it should be mentioned that these errors also depend significantly on the rotation rate considered. For instance, the relative difference $\left(\tau_{\text{r}}^{\text{GR}} - \tau_{\text{r}}^{\text{newt}} \right) /\tau_{\text{r}}^{\text{GR}}$  obtained for a 1.4~M$_{\odot}$ neutron star varies from $\sim 30\%$ at 10~Hz to $\sim -10\%$ at 327~Hz, using EoS~I.

  \subsection{Astrophysical considerations }
  \label{astro}
    
\begin{figure}
\includegraphics[width = 0.49\textwidth]{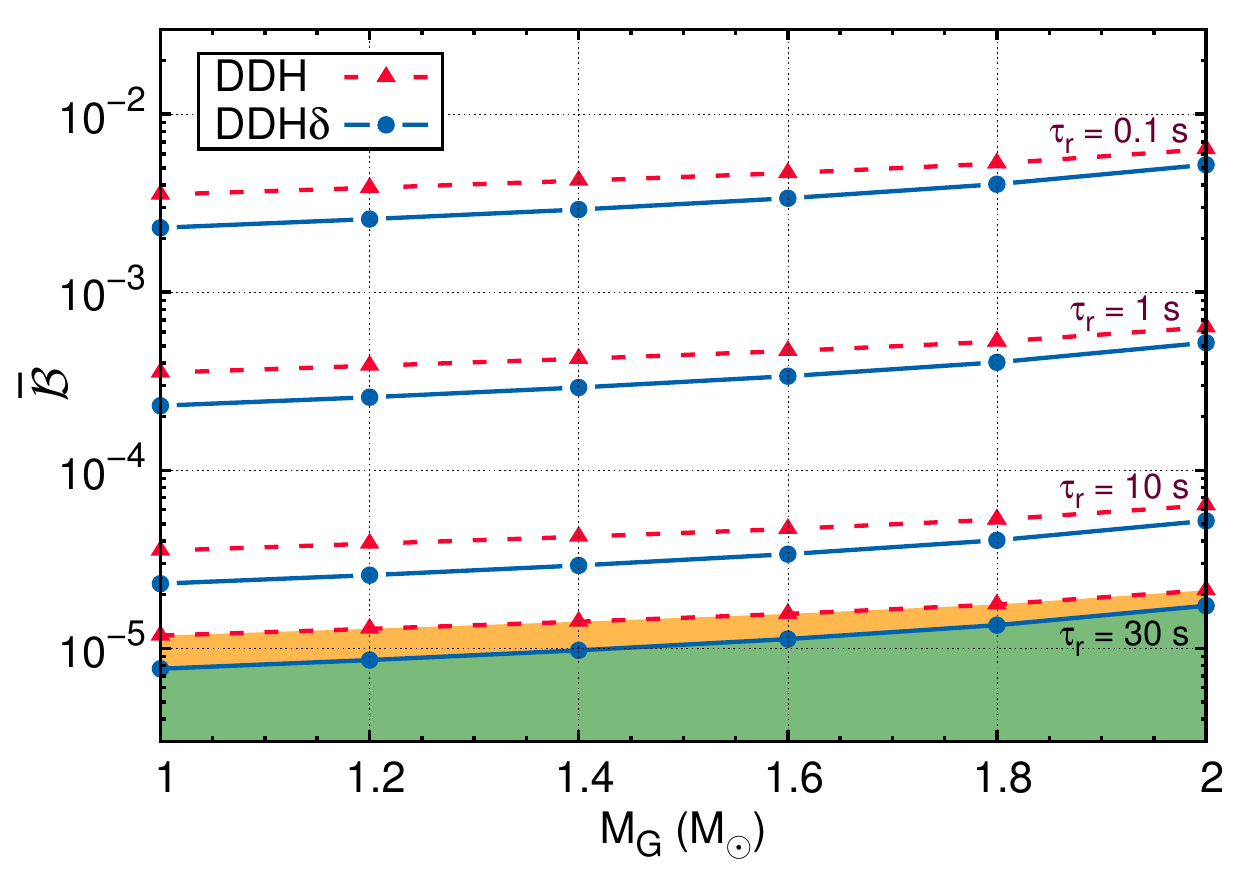}
\caption{Lines of constant rise time $\tau_\text{r}$ in a $\bar{\mathcal{B}}$ - $M_{\text{G}}$ diagram, assuming $f = 11.19$ Hz and $\Delta\Omega/\Omega=10^{-6}$. Results obtained from the DDH($\delta$) EoS are shown in solid (dashed) lines. The $\tau_r = 30$ s line refers to the upper limit from \citet{dodson2007two}.}
\label{fig:vela}
\end{figure}
 
In Fig.~\ref{fig:vela}, lines of constant spin-up time $\tau_{\text{r}}$ are displayed in the $M_\text{G}$ - $\bar{\mathcal{B}}$ plane, using DDH and DDH$\delta$ EoSs. These results are plotted for $f= 11.19$ Hz, which corresponds to the Vela pulsar. Considering the current upper limit $\tau_{\text{r}}~<~30$~s \citep{dodson2007two}, the mutual friction parameter $\bar{\mathcal{B}}$ should be higher than $\sim 10^{-5}$ to explain Vela glitches. Using Eq.~(\ref{drag_to_lift}) with $\bar{\mathcal{B}}\approx \mathcal{B}$, this limit implies that the averaged drag-to-lift ratio $\mathcal{R}$ should verify $10^{-5} \lesssim \mathcal{R} \lesssim 10^5$, which is not very constraining regarding the diversity of dissipative mechanisms that could give rise to mutual friction and the corresponding microscopic uncertainties. 
Note that similar conclusions were reached by \cite{glampedakis2009superfluid} from the analysis of post-glitch relaxation data. 
Nevertheless, as the dependence of the spin-up time on $M_\text{G}$ is much less pronounced than on $\bar{\mathcal{B}}$, future more stringent observational limits on $\tau_\text{r}$  shall put interesting constraints on the process governing the angular momentum transfer during the spin up.

\begin{figure}
\includegraphics[width = 0.49\textwidth]{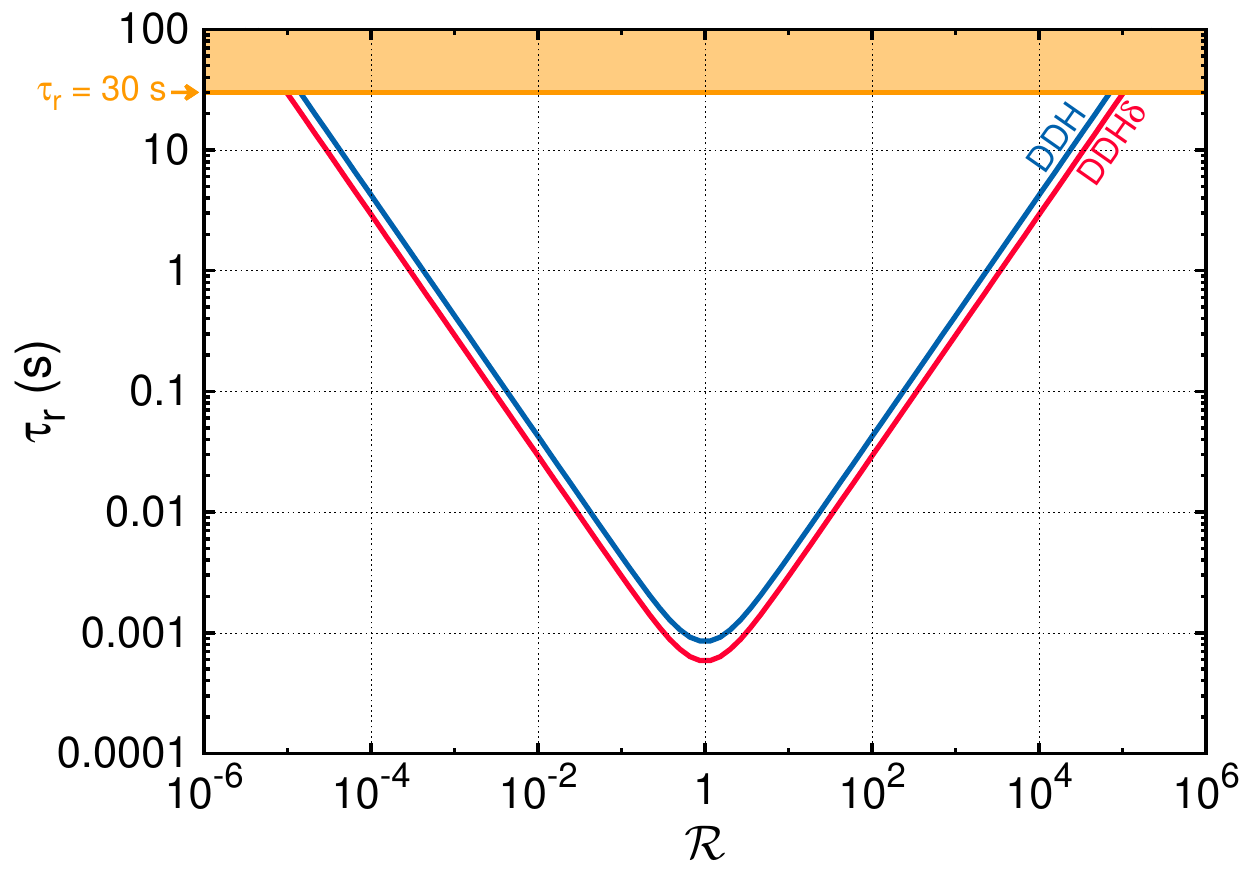}
\caption{Glitch rise time $\tau_{\text{r}}$ as a function of the drag-to-lift ratio $\mathcal{R}$, for a star spinning at 11.19 Hz with $M_{\text{G}} = 1.4$ M$_{\odot}$. The glitch amplitude considered here is $\Delta \Omega / \Omega = 10^{-6}$. Results obtained from the DDH($\delta$) are shown in blue (red). The 30 s upper limit is also displayed. 
}
\label{fig:tau_r_R}
\end{figure}

Finally, since $\bar{\mathcal{B}}\leq 1/2$, the shortest possible rise time (\ref{lower_bound}) associated with Vela glitches, $\tau_{\text{r}} \simeq 0.6-0.8$~ms for a 1.4 M$_{\odot}$ star (see Fig.~\ref{fig:tau_r_R}), is found to be comparable with the hydrodynamical time scale. Nevertheless, the actual value of  $\tau_{\text{r}}$ is presumably much longer in view of current estimates of the mutual friction parameters. Therefore, the whole dynamical evolution of star during the spin up can be accurately computed by considering a sequence of stationary configurations only.

\section{Gravitational waves}
\label{gws}

In this section, we study the amount of gravitational waves emitted through the time evolution of the mass quadrupole of the star, as a consequence of the changes in the fluid angular velocities during the spin up. Because of the small deviation from spherical symmetry, the variations of the mass quadrupole are expected to be very small. 
Thus, we focus on the DDH EoS, with which the equilibrium configurations obtained are sufficiently accurate to highlight these tiny variations (see \cite{sourie2016numerical}).

\subsection{Mass quadrupole}
\label{quadrupole}

To compute the coordinate-independent mass quadrupole moment $Q$ of the star, we follow the prescription given by Eq.~(11) of \cite{pappas2012revising} - see also \cite{friedman2013rotating}. The sign of the mass quadrupole moment is chosen such that $Q<0$ for an oblate spheroid. 
Using the DDH EoS, $|Q|$ decreases as time evolves, which means that the shape of the star is getting less and less oblate. 

For realistic glitch amplitudes, numerical results show that the time evolution of the mass quadrupole can be very well approximated by the formula
\begin{equation}
\label{fit_q}
 Q(t) = \Delta Q \left(1-\exp(-t/\tau_{\text{r}})\right) + Q_0,
 \end{equation}
where $\tau_\text{r}$ corresponds to the spin-up time scale discussed in the previous sections.
All other input parameters fixed, the variation $\Delta Q$ of the quadrupole moment during the glitch event is found to verify $\Delta Q \propto \Omega^2 \times \Delta \Omega / \Omega$. Whereas the dynamical evolution of $Q$ depends indeed on the mutual friction parameter through $\tau_{\text{r}}$, the variation $\Delta Q$ does not depend on $\bar{\mathcal{B}}$ because this latter does not play any role in determining the initial and final equilibrium configurations. On the other hand, the assumption on chemical equilibrium (Sec.~\ref{chemical}) affects $\Delta Q$: assuming $\Delta \Omega / \Omega = 10^{-4}$ and $M_{\text{G}} = 1.4$ M$_{\odot}$, the relative differences on $\Delta Q$ between cases (i) and (ii) are of the order of $\sim 1 \%$ for $f= 65$ Hz and $\sim 15 \%$ for $f= 327$ Hz.

\subsection{Gravitational wave amplitudes}

At leading order in a multipolar expansion, the gravitational radiation field is given by the so-called quadrupole formula, see e.g. \cite{bonazzola1996gravitational, letiec2016theory}. Since the star remains axisymmetric during the spin up, the $\times$ polarization of the wave strain vanishes. On the other hand, the $+$ polarization reads  
\begin{equation}
\label{h+}
h_+(t) = - \frac{3}{2} \frac{G}{Dc^4}\sin^2 i \times \ddot{Q}\left(t-D/c\right), 
\end{equation}
where $D$ is the distance to the pulsar and $i$ denotes the angle between the rotation axis of the star and the direction from the star's centre to the Earth. Using (\ref{fit_q}), the gravitational wave amplitude (\ref{h+}) is given by 
\begin{equation}
\label{h+_t}
h_+(t) = h_0 \sin^2 i\times \exp\left(-(t-D/c)/\tau_\text{r}\right), 
\end{equation}
where the characteristic amplitude $h_0$ is defined as
\begin{equation}
\label{def_h0}
h_0 = \frac{3}{2}\frac{G}{Dc^4}\frac{\Delta Q}{\tau_r^2}.
\end{equation}

In the frequency domain, the corresponding characteristic strain $h_c(f_{\text{GW}})$ is given from the Fourier transform $\tilde{h}(f_{\text{GW}})$ of the signal $h_+(t)$ through the relation
\begin{equation}
h_c(f_{\text{GW}}) = 2 f_{\text{GW}} \times |\tilde{h}(f_{\text{GW}})|,
\end{equation}
see, e.g., \cite{moore2015gravitational}. Using (\ref{h+_t}), the characteristic strain reads 
\begin{equation}
\label{strain}
h_c(f_{\text{GW}}) = \frac{h_0 \sin^2 i }{\pi} \times \frac{f_{\text{GW}}/f_0}{\sqrt{1+(f_{\text{GW}}/f_0)^2}}, 
\end{equation} 
where we have introduced the characteristic frequency
\begin{equation}
\label{def_f0}
f_0 = \frac{1}{2\pi \tau_{\text{r}}}.
\end{equation}

Assuming $M_{\text{G}} = 1.4$ M$_{\odot}$ and beta equilibrium at the center of the star, the gravitational wave characteristic amplitude (\ref{def_h0}) and frequency (\ref{def_f0}) obtained from the DDH EoS can be well-fitted by the following expressions
\begin{eqnarray}
\label{estim_ho}
h_0  \simeq &&1.0 \times 10^{-37} \Bigg(\frac{D}{1 \text{\ kpc}}\Bigg)^{-1} \Bigg(\frac{\bar{\mathcal{B}}}{10^{-3}}\Bigg)^2 \nonumber \\
 && \times \Bigg(\frac{\Omega}{10^2 \text{\ rad.s}^{-1}}\Bigg)^4\Bigg(\frac{\Delta \Omega/ \Omega}{10^{-6}} \Bigg), 
\end{eqnarray}

and 
 \begin{equation}
 \label{estim_fo}
f_0  \simeq 0.535 \Bigg(\frac{\bar{\mathcal{B}}}{10^{-3}}\Bigg) \Bigg(\frac{\Omega}{10^2 \text{\ rad.s}^{-1}}\Bigg) \text{\ Hz},
 \end{equation}
provided that $\Delta \Omega /\Omega \ll 1$. 
For low rotation frequencies, typically $f \lesssim 65$ Hz, the two expressions (\ref{estim_ho}) and (\ref{estim_fo}) approximate $h_0$ and $f_0$ with a precision better than a few percent and $\sim 0.1 \%$ respectively. For a star spinning at 327 Hz, the precision of these estimates is reduced to $\sim 10 \%$. Note that the prefactors involved in (\ref{estim_ho}) and (\ref{estim_fo}) are smaller for higher gravitational masses, mainly because the rise times are longer: for instance, we get $3.7\times 10^{-38}$ and 0.357 for $M_{\text{G}} = 2$ M$_{\odot}$. 

It should be remarked here that the quadrupole formula (\ref{h+}) is only valid in the slow-motion approximation, meaning that the frequency $f_{\text{GW}}$ of the gravitational wave emitted should satisfy the condition $f_{\text{GW}}\times R \ll c$, where $R$ is the characteristic size of the emitter. Taking $R\simeq 10^4$~m for the radius of the star, this leads to $f_{\text{GW}} \ll 3$ kHz, which in view of (\ref{estim_fo}) is well verified for glitching pulsars. Furthermore, we can easily check that the energy lost by gravitational waves emission is completely negligible with respect to that associated with mutual friction, such that Eq.~(\ref{evol_eq}) is valid.

 Applying (\ref{estim_ho}) and (\ref{estim_fo}) to the Vela pulsar, for which $\Delta \Omega / \Omega= 10^{-6}$, $f = 11.19$ Hz and $D\simeq 287$ pc \citep{dodson2003vela}, the constraint on the mutual friction parameter $\bar{\mathcal{B}}$ discussed in Sec.~\ref{astro}, \textit{i.e.} $10^{-5} < \bar{\mathcal{B}} < 0.5$, leads to 
\begin{equation}
h_0\sim 10^{-41} - 10^{-32} \ \ \ \text{and} \ \ \ f_0\sim 4 \text{\  mHz - 200 Hz}.
\end{equation}
Although the peak frequency $f_0$ is thus predicted to lie in the sensitivity bands of Advanced LIGO and Advanced Virgo \citep{acernese2015Advanced, LIGO2015Advanced, abbott2016prospects}, the corresponding gravitational-wave
signal is too weak to be detectable with present detectors.
However, other mechanisms associated with glitches such as Ekman pumping, which we have not taken into account in this work, might lead to a much stronger gravitational wave signal \citep{vanEysden2008gravitational, bennett2010continuous}.

\section{Conclusion}
\label{conclusion}

In this paper, we have studied in detail the impact of general relativity on the global dynamics of giant pulsar glitches as observed in Vela. We have carried out numerical simulations of the spin up  triggered by the sudden unpinning of quantized vortices. To this end, we have computed the exchange of angular momentum between the  neutron superfluid in the core and the rest of the star within a two-fluid model including neutron-proton entrainment effects. Both fluids were assumed to be coupled by mutual friction arising from dissipative forces acting on individual vortices. Since the hydrodynamical time scale is typically much smaller than the glitch rise time, we have described the time evolution of the two fluids by a sequence of quasi-stationary axisymmetric rigidly rotating configurations following \cite{sourie2016numerical}. We have calculated the mutual friction torque considering straight vortices arranged on a regular array, following \cite{langlois1998differential}.

In order to get some physical insight, we first solved analytically the dynamical equations by expressing the change in the lag as $\delta \dot{\Omega} /\delta \Omega \approx -1/\tau_{\text{r}}$, where the characteristic spin-up time scale $\tau_{\text{r}}$ can be expressed in a form similar to that obtained  in the Newtonian limit (see, e.g., \cite{carter2001relativistic, sidery2010dynamics}). However, general relativity not only changes the structure of the star, but also impacts the fluid dynamics. In particular, frame-dragging effects induce additional fluid couplings of the same form as the entrainment arising solely from neutron-proton interactions. For all these reasons, general relativity can change substantially the glitch rise time.

To test the validity of this analytical approach and to assess the importance of general relativity, we have also solved numerically the equations governing the transfer of angular momenta. For this purpose, two different kinds of inputs are needed: macroscopic quantities (the rotation frequency of the star, the glitch amplitude and the neutron star mass) and microscopic properties (the EoS and the stellar-averaged mutual friction coupling $\bar{\mathcal{B}}$). We have explored in detail various stellar configurations, using two different relativistic mean-field EoSs and considering the observed properties of glitching pulsars. The results obtained by numerical simulations are found to be very well reproduced by the analytical approximation. In particular, the glitch rise time $\tau_\text{r}$ can thus be expressed in terms of the moments of inertia of the fluids, the stellar rotation rate and  $\bar{\mathcal{B}}$, which can be obtained from stationary configurations. Furthermore, we have studied the effects 
of general relativity on $\tau_\text{r}$ by using two 
different polytropic EoSs of the kind previously introduced by \cite{prix2005relativistic}. 
Both the effects of general relativity on the structure of the star and on the fluid couplings are found to be important and therefore realistic simulations of the global glitch dynamics should be carried out in full general relativity.
Depending on the stellar compactness and on the rotation rate, the
errors incurred by using Newtonian gravity instead of general relativity
are found to be very sensitive to the adopted EoS, and amount to $\sim 20-40 \%$. These errors, however, might not be the dominant source of uncertainties. In particular, neutron superfluid vortices may not be
arranged on a regular array parallel to the rotation axis, as assumed
here. The dynamics of superfluid vortices and proton flux tubes remain
highly uncertain, and warrant further studies.

Considering the current upper limit $\tau_{\text{r}} < 30$ s \citep{dodson2007two}, we have found that the mutual friction parameter $\bar{\mathcal{B}}$ should be higher than $\sim 10^{-5}$ to explain Vela glitches. Since $\bar{\mathcal{B}}$ represents the average over the whole star, the mutual friction coupling  $\mathcal{B}$ might be locally much stronger ($\mathcal{B}\sim 10^{-4}-10^{-3}$) as discussed for instance by \cite{sedrakian2005type} and \cite{haskell2014new}. In any case, since the actual value of  $\tau_{\text{r}}$ is found to be much longer than the hydrodynamical time scale for current estimates of the mutual friction forces, the whole dynamical evolution of star during the spin up can be accurately computed by considering a sequence of stationary configurations only. 

We have also determined the amount of gravitational radiation emitted by the star during the spin up. For this purpose, we have studied the time variation of the mass quadrupole moment of the star resulting from changes in the fluid angular velocities. Using the quadrupole formula, we have numerically computed the characteristic amplitudes and frequencies associated with glitch events. Although the peak frequencies are found to lie in the sensitivity bands of current interferometers like Advanced LIGO, the corresponding amplitudes are too small for the gravitational waves to be detected. Their observations would require to improve the sensitivity by orders of magnitude. In particular, the characteristic amplitude for Vela is estimated to be at most $\sim10^{-32}$ for the (unrealistic) value $\bar{\mathcal{B}}=1/2$. If existing, the most promising sources would thus be pulsars rotating much more rapidly than Vela and undergoing high amplitude glitches.

Although glitches are unlikely to be detected through gravitational waves, the Low Frequency Array (LOFAR) radio telescope \citep{stappers2011observing} 
and the future Square Kilometer Array (SKA) \citep{watts2015probing} will be able to observe the spin up with unprecedented accuracy. It would thus lead to much more stringent constraints on the characteristic time $\tau_{\text{r}}$ and thereby on the underlying glitch mechanism. This calls for more realistic models of glitching pulsars including the crust magnetoelasticity and superfluidity (whose formalism has been already developed, see, e.g. \cite{carterchachoua2006,carter2006relativistic}), and accounting for the local dynamics of quantized vortices.

\section*{Acknowledgements}

We would like to thank Isma\"el Cognard for instructive discussions and Armen Sedrakian for interesting suggestions. This work has been partially funded by the ``Gravitation et physique fondamentale'' action of the Observatoire de Paris (France), the Fonds de la Recherche Scientifique - FNRS (Belgium) under grant n$^\circ$~CDR J.0187.16, the PHC Tournesol (n$^{\circ}$~35904ZJ) scientific cooperation program between France and Belgium, and the European COST action MP1304 ``NewCompstar''.




\bibliographystyle{mnras}
\bibliography{biblio} 



\appendix

\section{Angular momentum transfers in two-fluid Newtonian model}
\label{limit_newt}

In Newtonian gravity, the fluid angular momenta  read
\begin{equation}
 \label{newto}
 \left\{
   \begin{array}{rcl}
J_{\n}  &=& I_{\n} \left(1 - \tilde{\varepsilon}_{\n}\right) \Omega_{\n} + I_{\n} \tilde{\varepsilon}_{\n} \Omega_{\p}, \\
J_{\p}  &=& I_{\p} \left(1 - \tilde{\varepsilon}_{\p}\right)\Omega_{\p} + I_{\p} \tilde{\varepsilon}_{\p} \Omega_{\n},
  \end{array}
\right. 
 \end{equation}
 see Appendix~A of \cite{sourie2016numerical}. The moments of inertia $I_{X}$ involved in (\ref{newto}) are given by the classical formula
 \begin{equation}
\label{mom_newt}
	I_{X} = \displaystyle \int_{\Sigma_{t}} \rho_{X}  r^2 \sin^2\theta \df^{\, 3}\! \Sigma_{\text{f}},
 \end{equation}
 where $\rho_{X}$ is the mass density of fluid $X$ and $\df^{\, 3}\! \Sigma_{\text{f}}$ stands for the volume element of flat spacetime, while $r$ and $\theta$ refer to the radial and polar coordinates respectively. The quantities $\tilde{\varepsilon}_X$, which characterize entrainment, are defined as
\begin{equation}
\label{mean_entr}
\tilde{\varepsilon}_X = \frac{\displaystyle \int_{\Sigma_{t}}
  \varepsilon_X \rho_X  r^2 \sin^2\theta \df^{\, 3}\!
  \Sigma_{\text{f}}}{\displaystyle \int_{\Sigma_{t}}  \rho_X  r^2
  \sin^2\theta \df^{\, 3}\! \Sigma_{\text{f}}},  
 \end{equation}
where the entrainment parameter $\varepsilon_{X}$ is given by
\begin{equation}
\label{eps_p_newt}
\varepsilon_{X} = \frac{2\alpha}{\rho_X},
\end{equation}
and $\alpha$ is defined as in \cite{prix2005relativistic}. The mean entrainment parameters are related through the relation $I_{\n} \tilde{\varepsilon}_{\n} = I_{\p} \tilde{\varepsilon}_{\p}$. More details can be found in Appendix~A of \citet{sourie2016numerical}.

\subsection{Mutual friction torque}
\label{torque_newt}

In the Newtonian limit, the geometric term $\chi_{\perp}^2$ (\ref{def_hperp}) appearing in Eq.~(\ref{mom2}) is simply given by 
\begin{equation}
\chi_{\perp}^2= r^2\sin^2 \theta.
\end{equation}
Moreover, in case of  \textit{constant} entrainment parameters throughout the star, the non-relativistic neutron vorticity $\varpi_{\n}$ (\ref{def_vort}) reads
\begin{equation}
\label{varpi_newt}
 \varpi_{\n} =  2 m_{\n}\left[ \Omega_{\n} + \varepsilon_{\n} \left(\Omega_{\p} - \Omega_{\n} \right)\right],
\end{equation}
where $m_{\n}$ is the neutron mass. Considering that in the non-relativistic limit the volume element $\df^{\, 3}\! \Sigma$ is $\df^{\, 3}\! \Sigma_{\text{f}} = r^2\sin \theta \df r \df \theta \df \varphi$ and $\Gamma_{\n}= 1$, the mutual friction torque (\ref{mom2}) thus reduces to
\begin{equation}
\label{gamma_newtonien}
\Gamma_{\text{mf}} ^{\text{newt}} = - \bar{\mathcal{B}}  I_{\n}  \omega_{\n}  \delta \Omega, 
\end{equation}
where the superfluid vorticity per unit mass $\omega_{\n}$ reads 
\begin{equation}
\omega_{\n} = \frac{\varpi_{\n}}{m_{\n}}= 2  \left[ \Omega_{\n} +   \varepsilon_{\n} \left(\Omega_{\p} - \Omega_{\n} \right) \right].
\end{equation}
The Newtonian limit (\ref{gamma_newtonien}) corresponds to Eq.~(58) of \citet{sidery2010dynamics}. The  quantity $\kappa$ introduced in Eq.~(\ref{def_kappa}) is given by 
\begin{equation}
\kappa = I_{\n}  \omega_{\n}.
\end{equation}
To a very good approximation, the difference between $\Omega_{\n}$ and $\Omega_{\p}$ can be neglected, so that $\omega_{\n} \approx~2\Omega_{\n}$ and therefore
\begin{equation}
\label{kappa_newt}
\kappa \approx 2 I_{\n} \Omega_{\n}.
\end{equation}

\subsection{Glitch rise time}
\label{rise_newt}

Using Eqs.~(\ref{newto}), the  moments of inertia defined by (\ref{def_I}) lead in the Newtonian limit to 
 \begin{equation} 
 \label{IXX_newt}
   \begin{array}{rcl}
   I_{\n\hspace*{-0.05 cm}\n} &=& I_{\n} \left(1 - \tilde{\varepsilon}_{\n}\right)  + \Omega_{\p} \displaystyle\frac{\partial I_{\n}}{\partial \Omega_{\n}}, \vspace*{0.15 cm} \\ 
      I_{\n\hspace*{-0.05 cm}\p} &=& I_{\n} \tilde{\varepsilon}_{\n}  + \Omega_{\p} \displaystyle\frac{\partial I_{\n}}{\partial \Omega_{\p}}, \vspace*{0.15 cm} \\
I_{\p\hspace*{-0.05 cm}\n} &=& I_{\p} \tilde{\varepsilon}_{\p}  + \Omega_{\p} \displaystyle\frac{\partial I_{\p}}{\partial \Omega_{\n}}, \vspace*{0.15 cm} \\
I_{\p\hspace*{-0.05 cm}\p} &=& I_{\p} \left(1 - \tilde{\varepsilon}_{\p}\right)  + \Omega_{\p} \displaystyle\frac{\partial I_{\p}}{\partial \Omega_{\p}},  \end{array}
 \end{equation}
where the partial derivatives have been evaluated for vanishing lag. For slowly rotating stars, $I_{\n}$ and $I_{\p}$ are approximately independent of the rotation rates and the Newtonian moments of inertia should thus verify $\hat{I}_{\n} \simeq I_{\n}$, $\hat{I}_{\p} \simeq I_{\p}$, $\hat{I} \simeq I_{\n}+ I_{\p} \equiv I$ and $ I_{\n\hspace*{-0.05 cm}\p} \simeq I_{\n} \tilde{\varepsilon}_{\n} = I_{\p} \tilde{\varepsilon}_{\p}$. Consequently, comparing (\ref{zeta}) and (\ref{kappa_newt}) yields $\zeta =1$ in Newtonian gravity. As a result, the glitch rise time (\ref{tau_r}) in the Newtonian limit is given by 
\begin{equation}
\label{tau_r_newt}
\tau_{\text{r}} =  \frac{I_{\p}}{I}\times\dfrac{\left(1-\tilde{\varepsilon}_{\p} - \tilde{\varepsilon}_{\n}\right) }{2\bar{\mathcal{B}}\Omega_{\n}},
\end{equation}
which coincides with Eq.~(69) from \citet{sidery2010dynamics}.

\subsection{Fluid couplings}
\label{I_newt}

 \begin{figure}
\includegraphics[width = 0.49\textwidth]{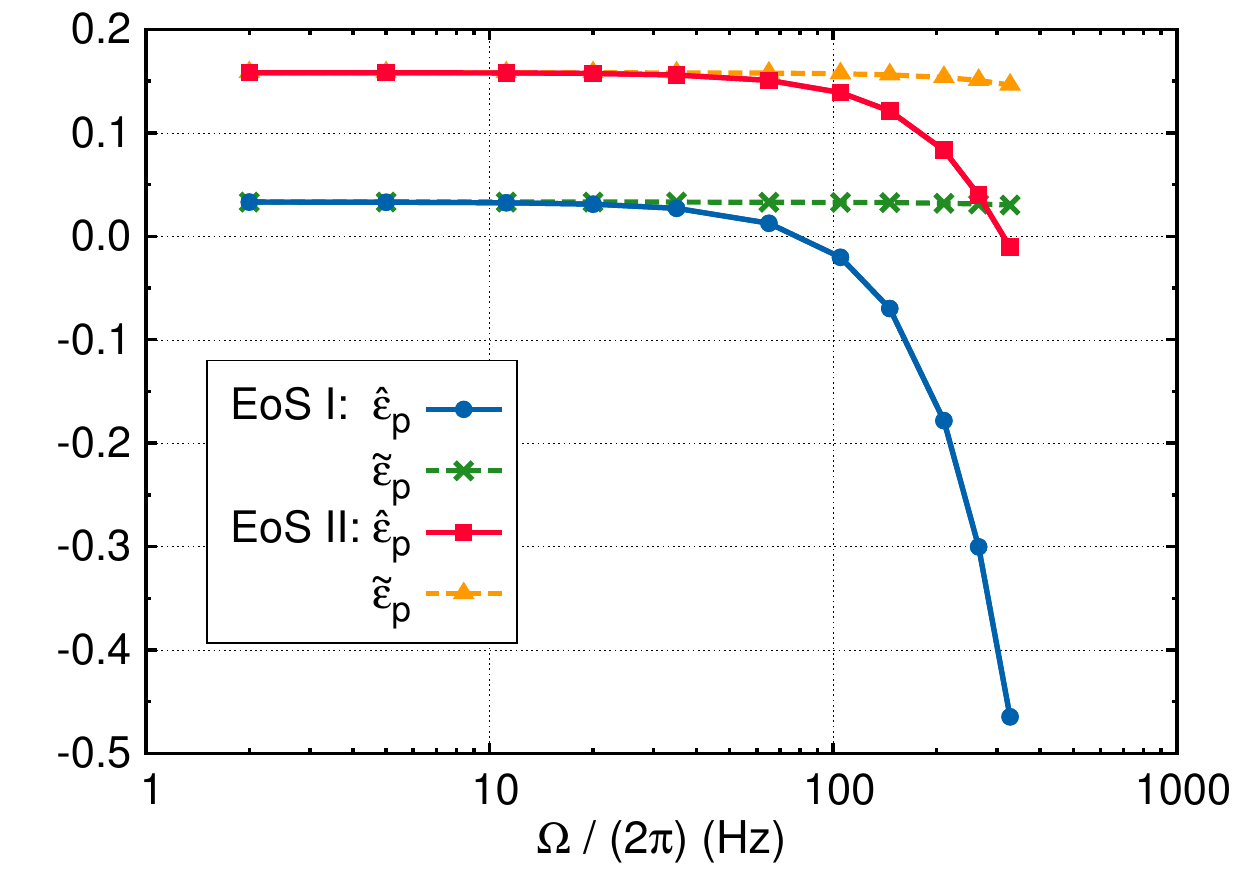}
   \caption{Coupling coefficient $\hat{\varepsilon}_{\p}$ and entrainment parameter $\tilde{\varepsilon}_{\p}$ as functions of the rotation frequency $f=\Omega/(2\pi)$ for a 1.4~M$_{\odot}$ neutron star, assuming corotation and beta equilibrium. The two EoSs considered are Newtonian versions of the EoSs used in Section~\ref{GR}.}
   \label{fig:Inp_newt}
 \end{figure}

Using Eqs.~(\ref{IXX_newt}), the Newtonian proton coupling coefficient $\hat{\varepsilon}_{\p}$ (\ref{eps_bar}) is given by
 \begin{equation}
 \label{eps_bar_newt}
 \hat{\varepsilon}_{\p}= \frac{I_{\n\hspace*{-0.05 cm}\p}}{\hat{I}_{\p}}=\frac{\tilde{\varepsilon}_{\p} + \displaystyle\frac{\Omega_{\p}}{I_{\p}}\displaystyle\frac{\partial I_{\p}}{\partial \Omega_{\n}}}{1 + \displaystyle\frac{\Omega_{\p}}{I_{\p}}\left(\displaystyle\frac{\partial I_{\p}}{\partial \Omega_{\p}} + \displaystyle\frac{\partial I_{\p}}{\partial \Omega_{\n}}\right)}.
\end{equation}  

We have computed Newtonian equilibrium configurations as discussed in Sec.~\ref{GR}.  In Fig.~\ref{fig:Inp_newt}, the coefficients $\hat{\varepsilon}_{\p}$ and $\tilde{\varepsilon}_{\p}$ are plotted with respect to the rotation frequency $f$, for a 1.4 M$_{\odot}$ neutron star, assuming corotation and beta equilibrium. As expected,  at low angular velocities $\hat{\varepsilon}_{\p}\simeq \tilde{\varepsilon}_{\p}$ to a very good approximation since the moment of inertia $I_{\p}$ is nearly constant.

For frequencies higher than $\sim100$~Hz, the effects of rotation on the stellar structure become non-negligible, and are twofold. First, the proton entrainment parameter $\tilde{\varepsilon}_{\p}$ is slightly decreased because the central density is lowered. More importantly, the moments of inertia change thus leading to large deviations between $\hat{\varepsilon}_{\p}$ and $\tilde{\varepsilon}_{\p}$.

\section{Constraints on the partial moments of inertia}
\label{app:constraints_stab}

From the application of the action principle to the asymptotically flat stationary states of an axisymmetric star composed of two fluids in circular motion with rigid angular velocities, the change in the energy of the star, between two nearby states, is given by 
\begin{equation}
\label{delta_E}
\delta E = \sum_{X} \Omega_{X} \delta J_{X},
\end{equation}
for a fixed total baryon mass, see Eq.~(3.5) from \cite{carter1975application}. Using the definitions (\ref{def_I}) of the moments of inertia, the variation in the angular momenta simply reads
\begin{equation}
\delta J_X = \sum_{Y} I_{X\hspace*{-0.05 cm}Y} \delta \Omega_{Y}.
\end{equation}  
Recalling that $I_{X\hspace*{-0.05 cm}Y}  = I_{Y\hspace*{-0.05 cm}X} $, Eq.~(\ref{delta_E}) is thus given by
\begin{eqnarray}
\delta E &=&\sum_{X,Y} I_{X\hspace*{-0.05 cm}Y}\Omega_{X}\delta \Omega_{Y} \\
		&=& \frac{1}{2}\left( I_{\n\hspace*{-0.05 cm}\n} \delta\left(\Omega_{\n}^2\right) + 2 I_{\n\hspace*{-0.05 cm}\p}\delta \left(\Omega_{\n}\Omega_{\p}\right) + I_{\p\hspace*{-0.05 cm}\p} \delta\left(\Omega_{\p}^2\right) \right).~~~~~~
\end{eqnarray}
Considering very small rotation rates, the energy $E$ of a rotating state is therefore given by 
\begin{equation}
 E = E_0 + \frac{1}{2}\left( I_{\n\hspace*{-0.05 cm}\n} \Omega_{\n}^2 + 2 I_{\n\hspace*{-0.05 cm}\p}\Omega_{\n}\Omega_{\p} + I_{\p\hspace*{-0.05 cm}\p}\Omega_{\p}^2\right),
\end{equation}
where $E_0$ stands for the energy of the static configuration. Rewriting this equation as follows
\begin{equation}
 E -E_0  =  \frac{1}{2}I_{\n\hspace*{-0.05 cm}\n}\left(  \Omega_{\n}  + \frac{I_{\n\hspace*{-0.05 cm}\p}}{I_{\n\hspace*{-0.05 cm}\n}} \Omega_{\p}\right)^2  + \frac{1}{2} \left(  I_{\p\hspace*{-0.05 cm}\p}- \frac{I_{\n\hspace*{-0.05 cm}\p}^{\ 2}}{I_{\n\hspace*{-0.05 cm}\n}}\right)\Omega_{\p}^2 ,
\end{equation}
the stability of the static state implies that the right-hand side should be strictly positive, leading to 
\begin{equation}
I_{\n\hspace*{-0.05 cm}\n} > 0  \ \ \  \text{and} \ \ \ I_{\n\hspace*{-0.05 cm}\n}I_{\p\hspace*{-0.05 cm}\p}-I_{\n\hspace*{-0.05 cm}\p}^{\ 2}>0, 
\end{equation}
which in turn gives $I_{\p\hspace*{-0.05 cm}\p} > 0 $.

\section{Relativistic coupling parameters}
\label{mean_entr_par}

Correcting a typo in Eq.~(A1) of \cite{sourie2016numerical}, the angular momentum of a fluid, say $X$, reads   
\begin{eqnarray}
  J_{X} =  &  \displaystyle  \int_{\Sigma_{t}} & \hspace*{-0.3 cm} \left[ \Gamma_{X}^2 n_{X}  \mu^{X}  U_{X}   + 2\alpha\frac{\Gamma_X^2}{\Gamma_{\Delta}^2}\left(\frac{\Gamma_{Y}}{\Gamma_{\Delta}\Gamma_{X}} U_{Y} -  U_{X} \right) \right]  \nonumber \\
    &\times &  \hspace*{-0.3 cm} B r \sin \theta\df^{\, 3}\! \Sigma. 
    \label{angu_mom_detail} 
   \end{eqnarray}
In this expression, $n_X$ and $\mu^X$ are respectively the particle density and the chemical potential of the fluid, as measured in its rest frame. The norms $U_X$ and $U_Y$ of the physical velocities of the fluids with respect to the ZAMO are given by
\begin{equation}
U_X = \frac{B}{N}\left(\Omega_X - \omega\right) r \sin \theta, 
\end{equation}
where $B$, $N$ and $\omega$ are different potentials involved in the spacetime metric, see \cite{sourie2016numerical}. The volume element is given by $\df^{\, 3}\! \Sigma = A^2 B  r^2 \sin  \theta\,{\rmn{d}}r\,{\rmn{d}}\theta\,{\rmn{d}}\varphi $.  $\Gamma_X$, $\Gamma_Y$ and $\Gamma_{\Delta}$ are Lorentz factors associated with $U_X$, $U_Y$ and $\Delta$, the relative speed between the fluids. The quantity $\alpha$, which characterizes the presence of entrainment, is linked to the more common entrainment parameter $\varepsilon_X$ through
\begin{equation}
\varepsilon_X = \frac{2\alpha}{n_X\mu^X\Gamma_{\Delta}^2},
\label{eps_entr}
\end{equation}
see section III-B of \cite{sourie2016numerical}.

\begin{figure*}
\includegraphics[width = 0.49\textwidth]{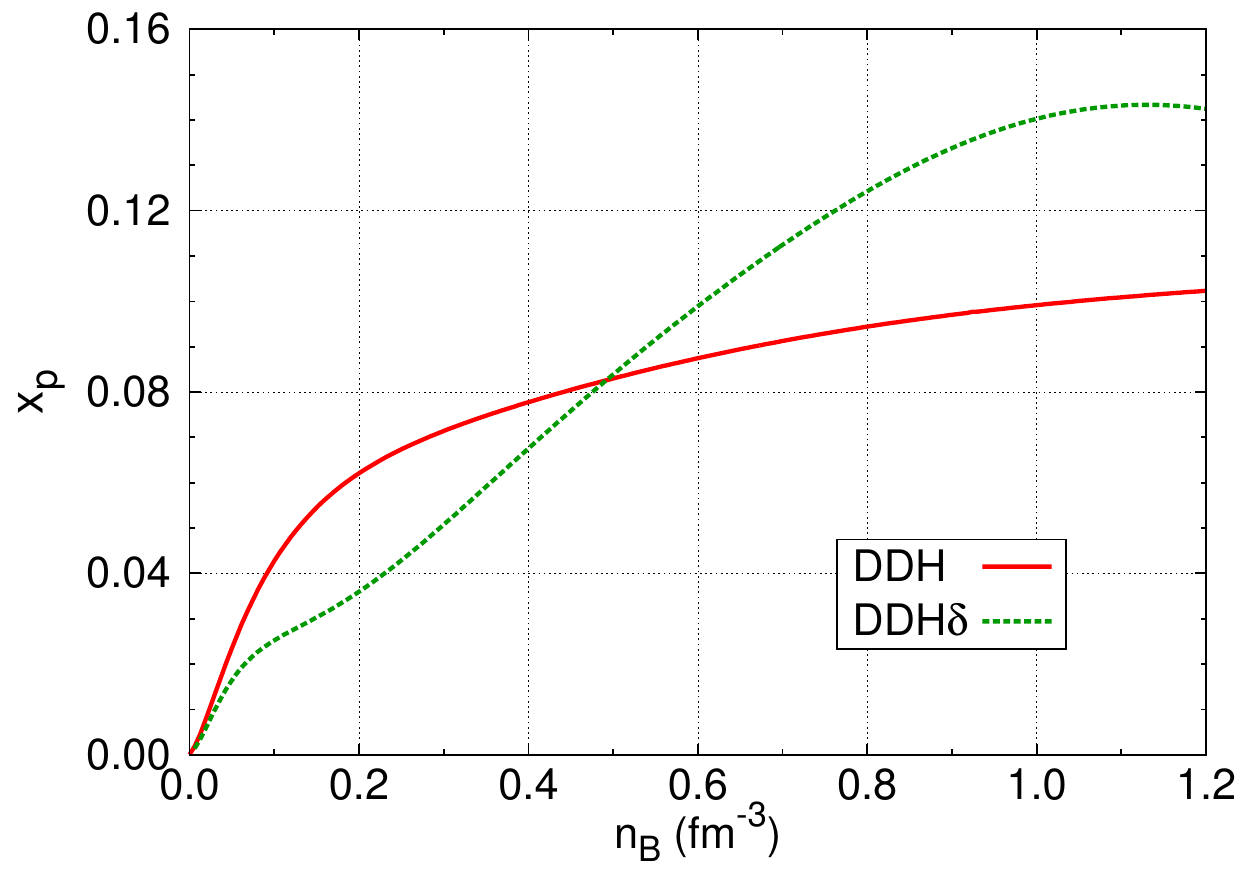}
\includegraphics[width = 0.49\textwidth]{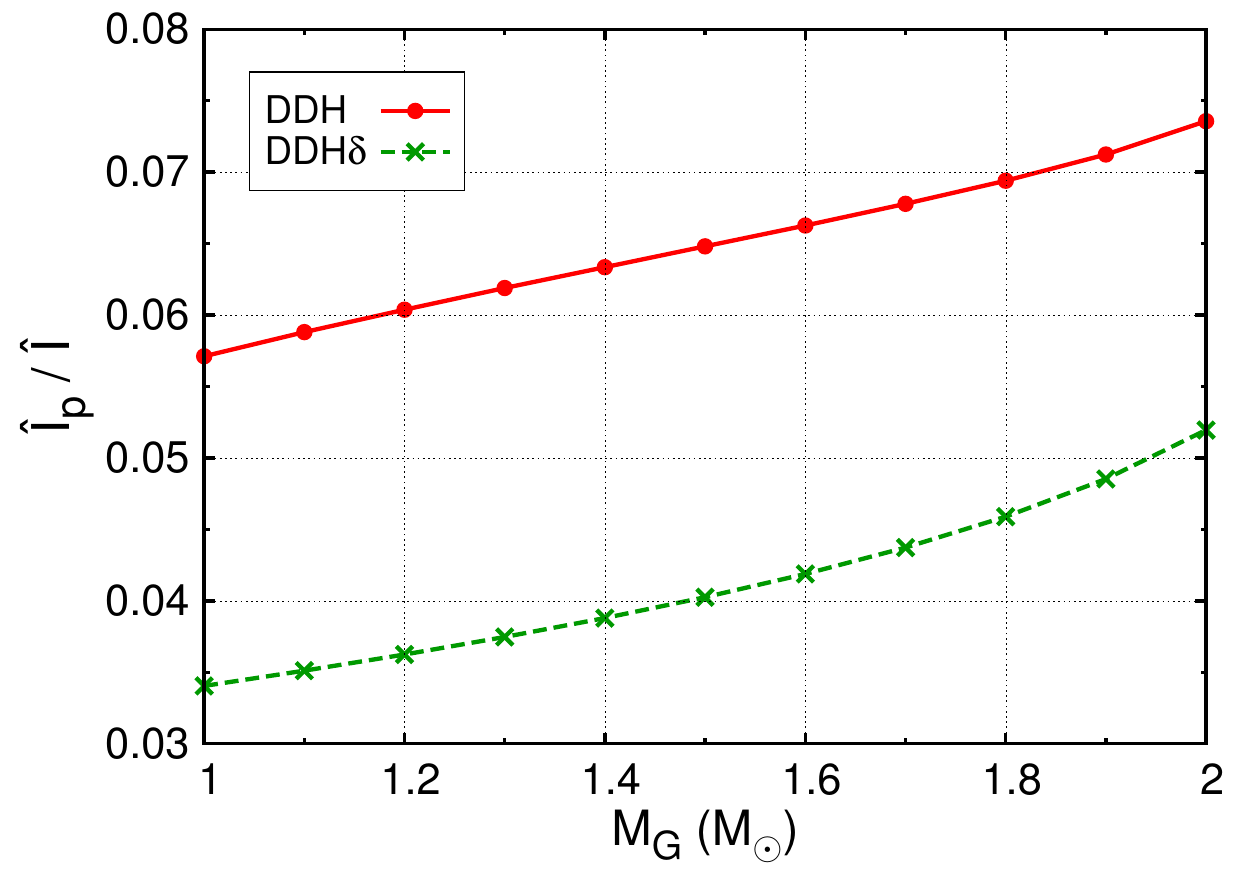}
   \caption{\textbf{Left:} Proton fraction $x_{\p}$ as a function of the total baryon density $n_{\text{B}} = n_{\n} + n_{\p}$, assuming corotation and beta equilibrium. \textbf{Right:} Ratio of the proton moment of inertia $\hat{I}_{\p}$ to the total one $\hat{I}$ with respect to the gravitational mass for a star spinning at 11.19~Hz, assuming beta equilibrium at the center. In both figures, results obtained from the DDH($\delta$) EoS are plotted with red solid (green dashed) lines.}
   \label{fig:xp}
 \end{figure*}

In the slow-rotation approximation $(\Omega_{\n}, \Omega_{\p} \ll \Omega_{\text{K}}$) and to first order in the lag $\delta \Omega = \Omega_{\n} - \Omega_{\p}$, Eq.~(\ref{angu_mom_detail}) becomes 
\begin{equation}
   \begin{array}{rcl}
   J_{X} &\approx& \displaystyle \int_{\Sigma_{t}}   n_{X}  \mu^{X}  \frac{B^2}{N} r^2 \sin^2 \theta \left(\Omega_X - \omega \right)\df^{\, 3}\! \Sigma \\[0.3 cm]
   &+& \displaystyle \int_{\Sigma_{t}}   n_{X}  \mu^{X}   \frac{B^2}{N} r^2 \sin^2 \theta  \ \varepsilon_X  \left(\Omega_Y - \Omega _X\right)\df^{\, 3}\! \Sigma,  \end{array}
   \label{J_corot}
\end{equation}
where the couplings by entrainment and Lense-Thirring effects are clearly visible, respectively through the terms $\varepsilon_X$ and $\omega$. 
 We now introduce the following quantities
\begin{equation}
\tilde{I}_X = \int_{\Sigma_{t}}   n_{X}  \mu^{X}  \frac{B^2}{N} r^2 \sin^2 \theta \df^{\, 3}\! \Sigma, 
\end{equation}
\begin{equation}
\tilde{\varepsilon}_X\tilde{I}_X = \int_{\Sigma_{t}}   n_{X}  \mu^{X}  \frac{B^2}{N} r^2 \sin^2 \theta \ \varepsilon_X \df^{\, 3}\! \Sigma, 
\label{mean_entra_par_RG}
\end{equation}
and
\begin{equation}
\tilde{\omega}_X\tilde{I}_X = \int_{\Sigma_{t}}   n_{X}  \mu^{X}  \frac{B^2}{N} r^2 \sin^2 \theta \ \omega \df^{\, 3}\! \Sigma, 
\label{mean_omega_RG}
\end{equation}
such that Eq.~(\ref{J_corot}) now reads 
\begin{equation}
J_X = \tilde{I}_X \left(\Omega_X - \tilde{\omega}_X \right)  + \tilde{\varepsilon}_X\tilde{I}_X \left(\Omega_Y- \Omega_X \right).
\label{JX_RG_coupling}
\end{equation}
In the Newtonian limit, $\tilde{I}_X$ and $\tilde{\varepsilon}_X$ are respectively given by Eqs.~(\ref{mom_newt}) and (\ref{mean_entr}) and $\tilde{\omega}_X$ simply vanishes, so that (\ref{JX_RG_coupling}) tends towards (\ref{newto}). We numerically find that $\tilde{\omega}_X$ can be approximated by a relation of the form
\begin{equation}
\tilde{\omega}_X = \varepsilon_{X\!X}^{\text{LT}} \ \Omega_X + \varepsilon^{\text{LT}}_{Y\!X}\  \Omega_Y,
\label{LT_coupling_terms}
\end{equation}
with a precision better than 0.1 $\%$ for a star spinning at 65~Hz or less. In this equation, $\varepsilon_{Y\!X}^{\text{LT}}$ represents the frame-dragging contribution of fluid $Y$ on fluid $X$ and $\varepsilon_{X\!X}^{\text{LT}}$ denotes the self-frame-dragging effect of fluid $X$. By making use of the different coupling parameters introduced so far, the angular momentum of fluid $X$ is given by 
\begin{equation}
J_X = \tilde{I}_X \left(1- \varepsilon_{X\!X}^{\text{LT}} - \tilde{\varepsilon}_X \right)\Omega_X  + \tilde{I}_X \left(\tilde{\varepsilon}_X- \varepsilon_{Y\!X}^{\text{LT}} \right)\Omega_Y.
\label{J_RG}
\end{equation}
To this level of approximation, the coupling parameter $\hat{\varepsilon}_X$ (\ref{eps_bar}) reads
\begin{equation}
\hat{\varepsilon}_X = \frac{\tilde{\varepsilon}_X- \varepsilon_{Y\!X}^{\text{LT}}}{1-\varepsilon_{Y\!X}^{\text{LT}}-\varepsilon_{X\!X}^{\text{LT}}}.
\label{eps_bar_th}
\end{equation}

\section{Realistic proton fractions}
\label{xp}

The proton fractions are plotted in left panel of Fig.~\ref{fig:xp} as functions of the total baryon density, for the DDH and DDH$\delta$ EoSs. For both EoSs, $x_{\p}$ is strongly increasing with the density, until $n_{\text{B}} \gtrsim 1.1$ fm$^{-3}$. The proton fraction profiles are very different between both EoSs because of the values considered for the symmetry energy and its slope at saturation density (see Table I of \citet{sourie2016numerical}). Consequently, as the mass of the stars increases, higher values of the proton fraction are reached and the quantity $\hat{I}_{\p} / \hat{I}$ increases, as can be seen in the right panel of Fig.~\ref{fig:xp}. This ratio is higher for the DDH EoS because, for densities found in neutron stars, the DDH proton fraction is more important than the DDH$\delta$ one and the central baryon densities are systematically higher in the case of DDH.


\bsp	
\label{lastpage}
\end{document}